\documentclass[twocolumn,showpacs,preprintnumbers,amsmath,amssymb, superscriptaddress]{revtex4-1}

\usepackage{graphicx}
\usepackage{dcolumn}
\usepackage{amsmath,mathrsfs,amssymb,amsthm}
\usepackage{hhline}
\usepackage{wasysym,enumerate,color}
\usepackage{epstopdf}
\usepackage{verbatim}
\usepackage{hyperref}
\usepackage{color}
\usepackage{ulem} 

\renewcommand{\emph}[1]{\textit{#1}} 

\definecolor{darkgreen}{rgb}{0,0.5,0}
\definecolor{purple}{rgb}{0.35,0,0.35}
\definecolor{orange}{rgb}{1,0.5,0}
\definecolor{darkred}{rgb}{.7,0,0}
\definecolor{darkblue}{rgb}{0,0,.3}
\definecolor{grey}{rgb}{.6,.6,.6}
\definecolor{dimgreen}{rgb}{0.2,0.6,0.1}
\hypersetup{colorlinks,linkcolor=blue,urlcolor=blue,citecolor=blue}

\newcommand{\be}{\begin{equation}}
\newcommand{\ee}{\end{equation}}
\newcommand{\bea}{\begin{eqnarray}}
\newcommand{\eea}{\end{eqnarray}}

\newcommand{\bR}{{\bf R}}
\newcommand{\bk}{{\bf k}}
\newcommand{\bq}{{\bf q}}

\newcommand{\g}{{\gamma}}

\newcommand{\cF}{{\cal F}}
\newcommand{\cT}{{\cal T}}
\newcommand{\cD}{{\cal D}}
\newcommand{\cJ}{{\cal J}}
\newcommand{\hn}{{\hat n}}
\newcommand{\hS}{{\hat S}}
\newcommand{\cS}{{\cal S}}
\newcommand{\cZ}{{\cal Z}}

\newcommand{\tg}{{\tilde g}}

\newcommand{\bra}[1]{\langle #1|}
\newcommand{\ket}[1]{|#1\rangle}

\newcommand{\w}{\omega}
\newcommand{\s}{\sigma}

\begin{document}
\title{Transmission of a microwave cavity coupled to localized Shiba states}
\author{Razvan Chirla}
\affiliation{Department of Physics, University of Oradea, 410087, Oradea, Romania}
\author{Andrei Manolescu}
\affiliation{School of Science and Engineering, 
Reykjavik University, Menntavegur 1, IS-101 Reykjavik, Iceland}
\author{C\u at\u alin Pa\c scu Moca}
\affiliation{BME-MTA Exotic Quantum Phase Group, Institute of Physics, Budapest University of Technology and Economics,
H-1521 Budapest, Hungary}
\affiliation{Department of Physics, University of Oradea, 410087, Oradea, Romania}

\date{\today}
\begin{abstract}

We consider a strongly correlated quantum dot, tunnel coupled to two superconducting leads 
and capacitively coupled to a single mode microwave cavity. When 
the superconducting gap is the largest energy scale, multiple Shiba states are formed inside 
the gap. The competition of these states for the ground state signals a quantum phase 
transition. We demonstrate that photonic measurements can be used to probe such 
localized Shiba states. Moreover, the quantum phase transition can be pinpointed exactly from the sudden change in the transmission signal. Calculations were performed using the numerical renormalization group approach. 

\end{abstract}


\maketitle

\section{Introduction} \label{sec:Introduction}
Quantum circuits combining elements of atomic physics, quantum optics, mesoscopic condensed matter  
physics and topological materials, could pave the way to the realization of fault-tolerant quantum  
computers~\cite{Xiang.13}. In such a hybrid circuit, a microwave coplanar waveguide resonator can act as a data bus between components of the setup. In particular, it has been shown experimentally that an artificial atom, such as a superconducting qubit \cite{Wallraff.04}, a quantum dot (QD)~\cite{Frey.11, Rundquist.11, Delbecq.11, Deng.15}, an Andreev dot \cite{Janvier.15}, or a double quantum dot~\cite{Frey.12, vanLoo.13, Kulkarni.14} coupled to a photon cavity, produces a measurable effect on the amplitude and phase of the transmitted electromagnetic field, as well as a shift and broadening of the resonant frequency of the cavity. The state-dependent frequency pull by the artificial atom can be used to entangle the state of the atom with that of the photons that pass through the resonator~\cite{Blais.04, Cottet.15}. By choosing an appropriate
driving frequency, high efficiency quantum non-demolition readout measurements of the state of the
atom can be performed, or alternatively, the state of the atom can be coherently controlled~\cite{Janvier.15}. 

So far, the coupling between mesoscopic  systems and microwave resonators has been studied by either neglecting the repulsive interaction between electrons~\cite{Skoldberg.08, Dmytruk.15a, LeHur.12, Dmytruk.15b}, by modeling the device as a   
two-level system~\cite{Kulkarni.14}, or employing various other approximations \cite{Schiro.14, Gudmundsson.12, Arnold.13}.

\begin{figure}[!tbhp]
\includegraphics[width=0.8\columnwidth]{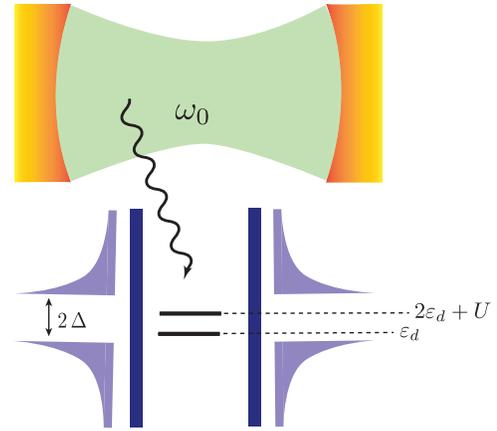}

\caption{(Color online) Sketch with the setup. An interacting quantum dot with on-site energy  $\varepsilon_d$ and Coulomb repulsion $U$, is coupled to two superconducting leads. The dot is capacitively coupled to a microwave cavity. The cavity is assumed to support a single mode with frequency $\omega_{0}$. 
The superconducting gap, $\Delta$ is the largest energy scale.
} 
\label{fig:sketch}
\end{figure}

In the present work we investigate  the response of a hybrid QD-resonator  device
by computing the phase and the amplitude of the transmitted signal as function of the incoming
microwave frequency. The system is displayed in Fig.~\ref{fig:sketch} and consists of a quantum dot
setup coupled capacitively to a microwave resonator. The dot is tunnel coupled to two superconducting leads. Such a system presents localizes states inside the superconducting gap -- also known as Shiba states~\cite{Shiba.68}. These are resonant states with a given 
parity~\cite{Yao.14}
(in our description they are either singlet or double states in the spin sector).
They are competing to form the ground state, implicitly driving the system through a 
quantum phase transition (QPT). On one side of the transition, 
when the ground state is a doublet, the photonic cavity is insensitive to the presence of the electronic system and
the transmission is unaffected, while on the other side of the transition,
when the ground state is a singlet, the Shiba states are expected
to be responsive when probed by the cavity~\cite{Janvier.15}. 
We indeed find that, by attaching the cavity to a pair of 
external transmission lines, both the amplitude and the phase difference
between the outgoing and incoming fields carry the fingerprint of the interaction between 
the cavity and the electronic system, and furthermore one obtains 
direct information on the energies of the Shiba states. 
Moreover, if the system is driven across the QPT by tuning a control parameter,
such as $\varepsilon_{d}$ or $\Gamma$, 
we expect a sudden change in the transmitted signal as some of the transitions between 
the Shiba states become
active. In this way, the presence and the position of the quantum critical point (QCP) can be accurately detected.

\section{System At Equilibrium}

\subsection{Model Hamiltonian}\label{sec:Hamiltonian}
Our system consists of a quantum dot that is coupled to two superconducting leads at equilibrium. 
The dot is also capacitively coupled to a microwave cavity (see Fig.~\ref{fig:sketch}).
The setup is described in terms of the Anderson model, which by now emerges as a standard approach to study electron transport in such nanodevices. In this framework, the system Hamiltonian, 
$H_{\rm sys}$, consists of several terms. The dot itself can be modeled by a spinful interacting level of energy $\varepsilon_d$, with a Coulomb interaction strength
$U$. The dot Hamiltonian  can be written in terms of the 
occupation $\hn = \hn_{\uparrow}+\hn_{\downarrow}$ as
\be
H_{\rm dot} = \varepsilon_d\, \hn + U\, \hn_{\uparrow}\,\hn_{\downarrow}\;, \label{eq:H_dot}
\ee
where $\hn_\s=d^\dagger_\s d_\s$ ($\s=\{\uparrow, \downarrow\}$), and $d^{\dagger}_\s$ are the creation operators on the QD.
The localized orbital hybridizes with the states in the two superconducting leads via the tunneling Hamiltonian 
\be
H_{\rm tun} = \sum_{\alpha=\{L,R\}}\sum_{\s}\left ( V_{\alpha} d^\dagger_\s\, \psi_{\alpha\s}(0)+\rm{H.C.}\right ).\label{eq:H_tun}
\ee
We assume a left-right symmetry, $V_{\alpha}=V$, where $\alpha=\{L,R\}$, and a 
tunneling rate $\Gamma=2\pi \varrho_0 V^2$. Here $\varrho_0=1/2D$ is the constant density of states of a conduction band. In what 
follows, $D=1$ will serve as the energy unit. $\psi_{\alpha\s}(\bR)$ is the field operator
for the conduction electrons, $\psi_{\alpha\s}(\bR) = \int\frac{d\bk}{(2\pi)^3} c_{\alpha \bk \s}e^{-i\bk \bR}$,
and $c_{\alpha \bk \s}$ are the annihilation operators for the conduction band electrons in lead $\alpha$, with momentum $\bk$ and spin $\s$. They 
satisfy the anticommutation relations: 
$\{c_{\alpha \bk \s}, c^{\dagger}_{\alpha' \bk' \s'} \} = (2\pi)^3 \delta(\bk -\bk')\delta_{\s\s'}\delta_{\alpha\alpha'}$.
The dot is also capacitively coupled to the cavity
\be
H_{\rm int} = g\,\hn \left (a+a^\dagger\right ),\label{eq:H_int}
\ee
with $g$ the strength of the light-matter coupling, and  $a^\dagger$  a bosonic operator describing the creation of a photon in the cavity.
In our approach we assume that the resonator supports a single monochromatic mode $\omega_0$, so that the cavity Hamiltonian reduces simply to 
\be
H_{\rm cav} = \hbar\, \omega_0\, a^\dagger a\;.\label{eq:H_cav}
\ee

Most of the theoretical approaches
neglect the back-action from the cavity on the electronic states, and restore a finite coupling
when investigating the dynamics of the photons only. This is a well suited approximation for the tunneling
junctions~\cite{Mendes.15}  or QDs~\cite{Cottet.15,Dmytruk.15a} coupled
to normal leads as long as the inter-level energy spacing
of the electronic system $\delta_{l} \ll \hbar\, \omega_{0}$, otherwise the decoupling of the photons from the 
QD is no longer possible, and the electronic transport is
affected \cite{Gudmundsson.12,Arnold.13}. 
This condition is not satisfied in our setup, as the energy of the excited Shiba states inside the 
superconding gap, 
$E_{\rm S}\sim \hbar\,\omega_{0}$, so decoupling the photons from the QD is not possible. 
The resonant coupling
between the cavity and the QD  is evident from the avoided level crossing between the gap states
(see Fig.~\ref{fig:levels}(b) for details.) 

Finally, the metallic leads that couple to the dot are considered as BCS superconductors and are described by the Hamiltonian
\begin{multline}
H_{\rm leads} = \sum_{\alpha=\{L,R\}}\int \frac{d\bk}{(2\pi)^3}\sum_\s \xi_{\alpha\bk\s}c^\dagger_{\alpha \bk \s}c_{\alpha \bk \s}+\\
(\Delta_{\alpha} c^\dagger_{\alpha \bk \uparrow} c^\dagger_{\alpha -\bk \downarrow}+\mathrm{H.C.}). \label{eq:H_leads}
\end{multline}
In Eq.~\eqref{eq:H_leads}, $\Delta_\alpha=|\Delta|e^{-i\varphi_\alpha}$ stands for the superconducting gap in lead $\alpha$. For simplicity we use $\varphi_{\alpha}=0$ in what follows. We first study the ground state, excitation spectrum and the spectral properties of the system Hamiltonian
\be
H_{\rm sys} = H_{\rm dot }+H_{\rm cav}+H_{\rm tun}+ H_{\rm int} +H_{\rm leads},
\label{eq:H_sys}
\ee
by treating all the interactions in the setup 
on equal footing. We call this an equilibrium configuration in the sense that 
the cavity is not connected to the external transmission lines, but only to the quantum dot. As a  method we use the numerical renormalization group (NRG)~\cite{Wilson.75, Wilson.80}, in a manner similar to how it was done previously for the Anderson-Holstein model \cite{Hewson.02, Jeon.03}. Details of the NRG calculations are provided in Appendix \ref{app:NRG}. In our numerical calculations we consider only
the particle-hole symmetrical situation, i.e. $\varepsilon_d=-U/2$. Although
we are always at the particle-hole symmetrical point, we shall not perform the calculations in the 
Kondo limit \cite{hewson} (corresponding to $-\varepsilon_{d}=U/2\to \infty$), but allow for large quantum charge fluctuations. In our calculations,
the charging energy of the dot is alway smaller than the superconducting gap, i.e. $\Delta>\{U, |\varepsilon_{d}|\}$. 
This is a required condition for the second singlet-like excited Shiba state (labeled $S_{+}$ 
in Fig.~\ref{fig:levels} (a)) to survive inside the gap. In this way, the superconducting gap supports two bound states of the same parity, which allows photonic transitions between them~\cite{Janvier.15}. 

For a better understanding of the whole picture we shall also discuss shortly the physics of the 
electronic part of the system in the absence of the photons. 
It corresponds to the Hamiltonian
\be
H_{\rm el} = H_{\rm dot }+H_{\rm tun}+H_{\rm leads}.
\label{eq:H_el}
\ee
Although the NRG is a powerful and exact method, the results are not always intuitive enough,
and for that, we shall supplement our NRG results, where possible, with analytical descriptions. 

\subsection{Shiba states}
\subsubsection{Shiba states in the absence of photons}

\begin{figure}[!tbhp]
\includegraphics[width=0.85\columnwidth]{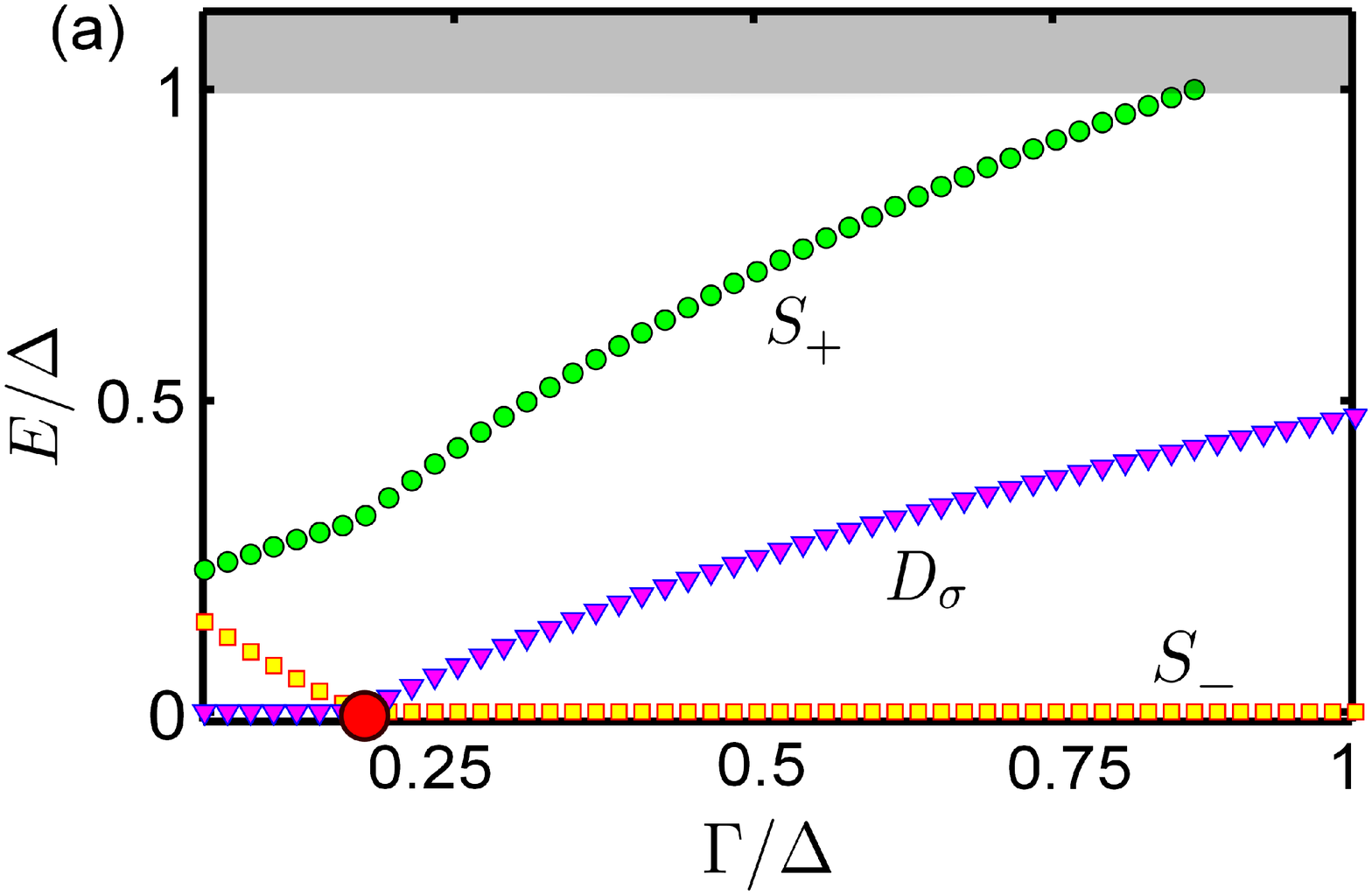}
\includegraphics[width=0.85\columnwidth]{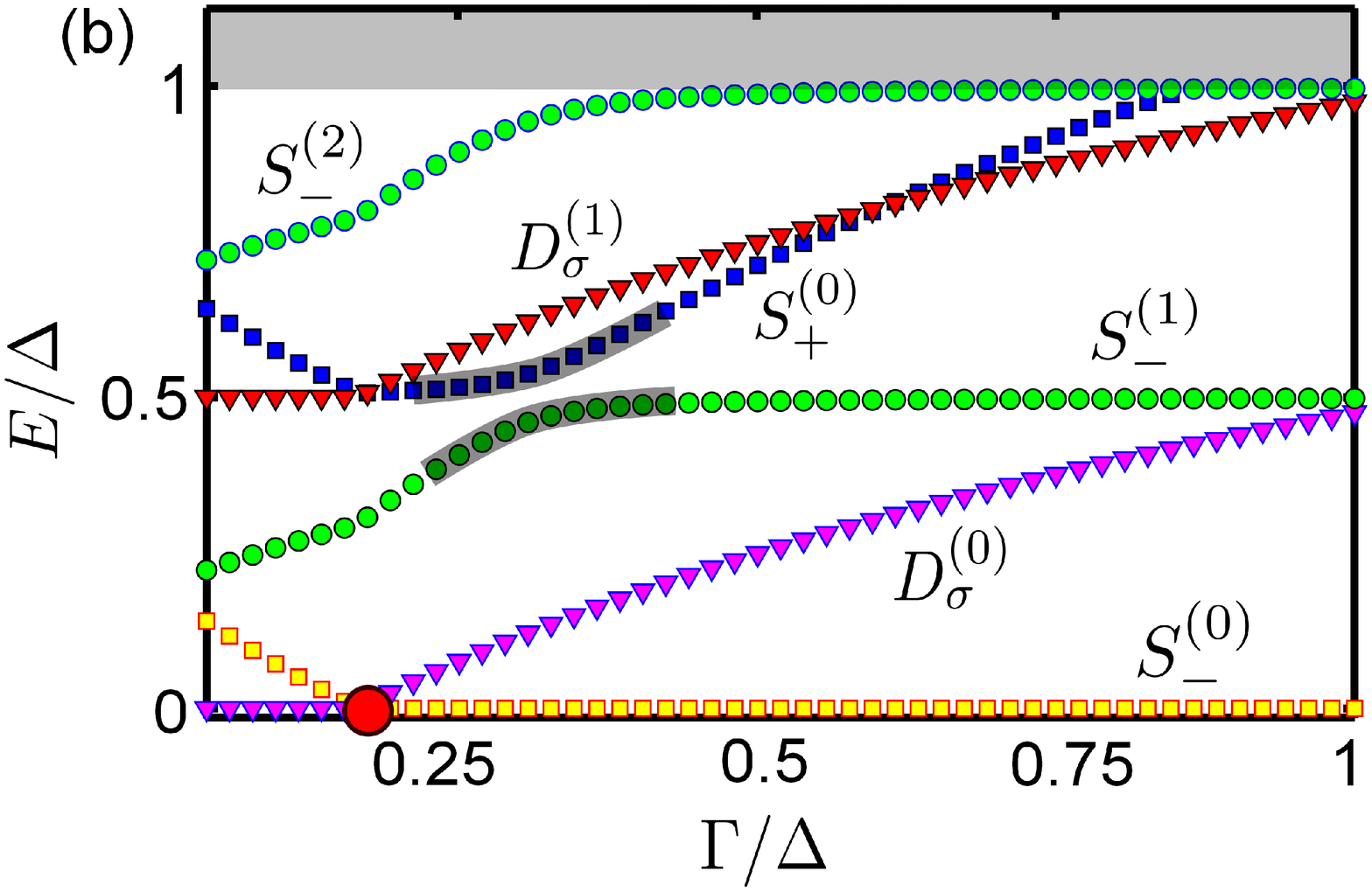}
\caption{(Color online) (a) Shiba states inside the gap in the absence of photons. The QCP is marked by the red dot. 
(b) Shiba states in the presence of the photons.  The other parameters are fixed to: $\omega_0/\Delta=0.5$, $g/\Delta=0.05$ and $U/\Delta=0.4$. 
For this set of parameter the quantum critical point corresponds to $x_{\rm QCP}=\left (\Gamma/\Delta\right)_{\rm QCP}\simeq 0.17$. The shaded lines emphasize the avoided level crossing. 
The shaded area above $E=\Delta$ represents the continuum. 
} 
\label{fig:levels}
\end{figure}

 So far, the Hamiltonian 
$H_{\rm el}$ given in Eq.~\eqref{eq:H_el} which does not include the coupling to the resonant cavity, has been studied 
extensively in the literature \cite{Shiba.68, Rozhkov.99, Zonda.15}. In the limit when 
charge fluctuations are frozen, $U\to \infty$, a pair of states develops in the 
gap: a singlet-like state 
(labeled $S_{-}$ in Fig.~\ref{fig:levels}) and a doublet $D_{\s}$.
Depending on the setup parameters, each of these states can be the ground state. Consequently,
the system develops a quantum phase transition. This transition can be understood as the competition between the superconducting correlations and the Kondo screening, and 
takes place when $\Delta\sim T_K$. Here $T_{K}$
is the Kondo temperature characterizing $H_{\rm el}$ when the external leads are 
in the normal state. 
On one side of the transition, when $\Delta>T_K$, the 
ground state is the doublet $D_{\s}$, as the local moment remains unscreened, while on the other side, $\Delta <T_K$, the Kondo screening
wins and the ground state becomes the many-body Kondo singlet $S_{-}$. 

If charge fluctuations are allowed, and if $U<\Delta$, a second singlet state $S_{+}$ develops inside the gap. 
In the non-interacting limit, when $U\to 0$,  
its energy is exactly $E_{S_{+}}= 2\,E_{D_{\s}}$~\cite{Hecht.08}. Increasing $U$ sufficiently, its energy starts to 
shift to larger values, and when $U\sim \Delta$, it merges with the continuum. 

As we are interested in the bound states inside the gap, it is natural to consider the $\Delta\to \infty$ limit. It was previously 
shown~\cite{Meng.09} that the QPT transition can be nicely captured in this limit, and that an effective model can be constructed 
by integrating out the leads:
\begin{multline}
H_{\rm el}^{(\Delta\to \infty)} = \sum_{\s}\xi_{d}\,d^\dagger_\s d_\s-\Gamma \left (d^\dagger_\uparrow d^\dagger_\downarrow +\mathrm{h.c.}\right)+\\
+{U\over 2} \left (\hn-1 \right )^2,\label{eq:H_el1}
\end{multline}
where $\xi_d= \varepsilon_d +U/2$. The Hamiltonian in Eq.~\eqref{eq:H_el1} can be diagonalized in a local basis as
\be
H_{\rm el}^{ (\Delta\to \infty)} = \sum_{\g = \pm, \s} E_\g \ket\g \bra \g\;.
\label{eq:H_el2}
\ee
With the notations $u=(1+\xi_d/2E_d)/2$ and $v=(1-\xi_d/2E_d)/2$, where $E_d=\sqrt{\xi_d^2 +\Gamma^2}$, the eigenstates
can be grouped into a pair of singlets $\ket \pm$:
$\ket + =  (u\, d^\dagger_\uparrow\, d^\dagger_\downarrow+v)\ket 0 $, 
$\ket - =  (-v^*\, d^\dagger_\uparrow\, d^\dagger_\downarrow+u)\ket 0$, 
and a doublet  $\ket \s= d^\dagger_{\s} \ket 0$. The corresponding energies are $E_{\pm} = \xi_d+U/2 \pm \sqrt{\xi_d^2+\Gamma^2}$ and 
$E_\s = \xi_d$. Since $E_- <E_+$, the  QPT is recovered  when the 
singlet $\ket -$ and the doublet
state $\ket \s $ are degenerate in energy: $E_- = E_\s$. 
Since $\Delta\to \infty$, all these states in Eq.~\eqref{eq:H_el2} are actually resonant 
levels that live inside the gap and are the so-called Shiba 
states~\cite{Shiba.68}. In this limit, there is a 
one to one correspondence with the states obtained in 
the NRG: $\ket \pm \leftrightarrow S_{\pm}$ and $\ket \s \leftrightarrow D_{\s}$. 

\subsubsection{Shiba states and the Jaynes-Cummings Hamiltonian}

In the presence of a cavity, these states start to hybridize with the photons. Since the operator $\hn$ entering Eq.~\eqref{eq:H_int} is charge conserving, $H_{\rm int}$ couples 
only states with the same parity. Moreover, in the absence of the external field, the doublet states are always degenerate in energy, 
so there is no dipole matrix element between them, and they remain decoupled from the photonic field. In that regard, the following discussion is  
suitable for the  singlet side of the QPT, where the singlet $\ket -$ is the ground state
and the singlet $\ket +$ is the excited state.
%
Computing the dipole matrix element $\bra+ \hn \ket - = -2u\, v$, and introducing 
the regular notations $\sigma^+ = \ket +\bra -$, $\sigma^- =  (\sigma^+)^\dagger$, and using the rotating wave approximation, 
the Hamiltonian $H_{\rm sys}$ in Eq.~\eqref{eq:H_sys} can be rewritten simply as a Jaynes-Cummings Hamiltonian~\cite{Cummings.63} in the singlet sector plus a 
non-interacting part in the doublet sector
\begin{multline}
H_{\rm sys}^{(\Delta\to \infty)} ={1\over 2} \hbar\,\omega_A\,\sigma^z  + \hbar\,\omega_0 a^\dagger a - \hbar\,\tg\left(a\,\sigma^+ +\sigma^- a^\dagger \right)+
\\
+\sum_{\s=\{\uparrow,\downarrow\}}E_\s \ket\s \bra \s,
\end{multline}
where we have introduced the notations $\hbar\omega_A = 2 E_d$ and $\hbar\tg = g (\Gamma/E_d)$. This two-level system coupled to a photonic field has been studied thoroughly by now~\cite{Mandel}, so we won't discuss it further here. 

\subsubsection{Energy spectrum} 

In this section we discuss in more detail the NRG results for the energy spectrum in the limit when $\Delta$ is the largest energy scale
in the problem, i.e. $\Delta \gg \{|\varepsilon_d|, U, \Gamma, \hbar\, \omega_0\}$. We 
shall ignore the continuum $E>\Delta$ and discuss only the localized states. 

We start by presenting in Fig.~\ref{fig:levels}(a) the evolution of the Shiba states for $H_{\rm el}$ as function of $x=\Gamma/\Delta$ in the absence of the photons. For $x\le x_{\rm QCP}$,
the ground state consists of a BCS superconductor plus a free local moment (doublet ground state $ D_\s$), 
while in the other limit, when $x>x_{\rm QCP}$, the local
spin screened by the conductions electrons forms a many-body singlet ground state ($ S_-$). The other singlet state $ S_+$ is always 
higher in energy and never competes for the ground state. 
The QCP corresponds 
to the degeneracy in energy of the levels $S_-$ and $D_\s$ (market by a red dot in Fig.~\ref{fig:levels}). 

In Fig.~\ref{fig:levels}(b) we show the same Shiba states in the presence of the photonic field. The spectrum was obtained by solving the Hamiltonian~\eqref{eq:H_sys} exactly, with 5 photons in the system. We have checked that by adding extra photons, the spectrum inside the gap 
remains the same, and only higher energy states in the continuum are affected. 
Throughout our calculations, the photon energy is fixed to $\omega_0=\Delta/2$. First of all, except for a small Stark shift towards the doublet sector as $g$ is further increased, 
the photons have a negligible impact on the position of the QCP. 
Moreover, as the doublet state does not hybridize with the photonic states 
(there is no dipole matrix element), its energy remain unaffected, 
so that we can only see the formation of equally separated energy states in this channel (ladder states,  labeled as $D_\s^{(0)}$, $D_\s^{(1)}$,$\dots$). 
On the other hand, the photonic states hybridize with electronic ones in the singlet channel, and
this is signaled by the formation of the avoided level crossing when $\omega_0\simeq E_{S_{+}}$. In Fig.~\ref{fig:levels}(b) this is highlighted
by the shaded lines at $x \simeq 0.3$. The labeling of the lines  emphasizes both to which electronic states they belong, and
the number of photons used to build the states in the singlet sector, far away from the resonance.

Although in Fig.~\ref{fig:levels}(b) we present the situation when the 
position of the avoided level crossing is on the singlet side of the QPT,
its position as function of $x$ can be tuned to the other side of the QPT by simply tuning some control parameter, such as $\omega_0$. Notably, in Ref.~\cite{Janvier.15}, this resonant coupling has been 
measured recently by using a circuit quantum electrodynamic setup coupled to an
Andreev qubit.

\subsection{Electronic spectral functions}

\begin{figure}[!tb]
\includegraphics[width=0.9\columnwidth]{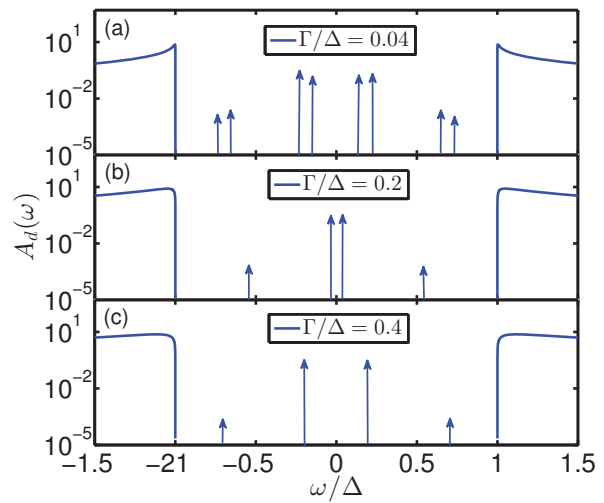}
\caption{(Color online) Spectral function for the on-site creation operator $d_{\s}^{\dagger}$ at T=0. The weights and positions for the transitions between the ground state and the excited Shiba states are indicated by vertical arrows. 
They correspond to the transitions between:  $D_{\s}^{0}\leftrightarrow \{S_{-}^{(0)}, S_{+}^{(0)}, S_{-}^{(1)},S_{-}^{(2)} \}$ for $\Gamma/\Delta =0.04$ when the ground state is $D_{\s}$, and to 
 $S_{-}^{(0)}\leftrightarrow \{D_{\s}^{(0)}, D_{\s}^{(1)}\}$ when $\Gamma/\Delta= \{0.2, 0.4\}$, when the ground state is $S_{-}$.
} 
\label{fig:1D_d}
\end{figure}

Here we discuss the results for the spectral function $A_{d}(\omega)$ of the operator
$d_{\sigma}^{\dagger}$:
\begin{equation}
A_{d}(\omega)= -{1\over \pi}\rm{Im}\, G_{d}(\omega)\, ,
\end{equation}
with $G_{d}(\omega)$ being the Fourier transform of the electronic Green's function:
$G_{d}(t)= -i\theta(t)\,\langle \{d_{\s}(t), d_{\s}^{\dagger}(0)\} \rangle$. 
Since $d^{\dagger}_{\s}$ is a charge $q=1$ operator, only the transitions between 
the ground state and the excited Shiba states, having different 
parity, are visible in the spectral density $A_{d}(\omega)$. 
For $x<x_{\rm QCP}$ (panel (a) in Fig.~\ref{fig:1D_d}) the ground state is 
the doublet $D_{\s}$.
In this case, the transitions $D_{\s}^{(0)}\to S_{\pm}^{(n)}, n=0,1,\dots$ are active.
For our set of parameters, there are four such singlet excited 
states inside the gap, implying 
four visible particle-like resonances for $\omega>0$. The associated pairs are the 
hole-like transitions at negative frequencies. In the singlet phase, the number 
of resonances reduces to half (see panels (b) and (c) in Fig.~\ref{fig:1D_d}), as the
transitions 
$S_{-}^{(0)}\to S_{\pm}^{(n)}$ become forbidden by symmetry.

\begin{figure}[!tb]
\includegraphics[width=0.9\columnwidth]{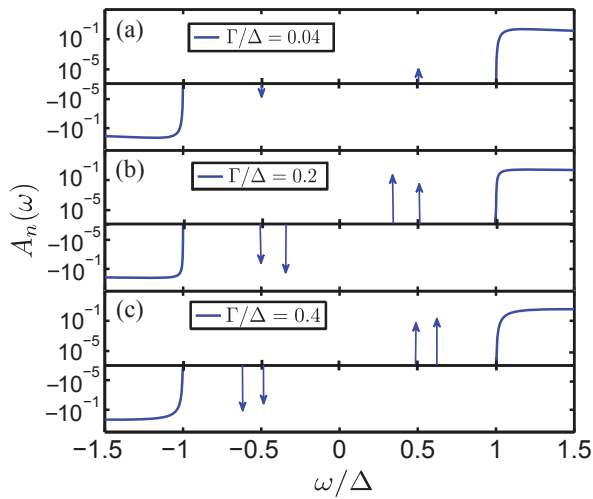}
\caption{(Color online) Bosonic spectral function for the occupation number operator of the dot, $\hn$ at T=0. The transitions inside the gap are between the ground state and 
excited states with the same parity.  
} 
\label{fig:1D_n}
\end{figure}

The observation of the Shiba states can be achieved by 
measuring the cavity transmission or reflection~\cite{Janvier.15}. In that regard, 
the change in the output signal gives information about the electronic susceptibility.
In Fig.~\ref{fig:1D_n} we display the spectral function for the dot occupation 
operator, $\hn=\hn_{\uparrow}+\hn_{\downarrow}$, that corresponds to the 
dissipative part of the electronic susceptibility. The transitions
between the states with the same parity are now visible. This is the reason why in the
singlet region, when $x>x_{\rm QCP}$, two transitions are visible, corresponding to 
$S_{-}^{(0)}\to \{S_{-}^{(1)}, S_{+}^{(0)}\}$. Therefore, for our 
specific setup, this quantity gives direct 
information on the existence of and transitions between various Shiba states. 
This will be discussed in more detail in Sec.~\ref{sec:transmission}.

\section{Microwave probed cavity}\label{sec:transmission}

So far we have discussed the system at equilibrium, but our main goal is to probe the Shiba states through photon
transport. For that, the resonator is coupled capacitively to two external transmission lines that transmit the input and output 
microwave signal. The total Hamiltonian takes the form
\begin{eqnarray}
H&=&H_{\rm sys}+\sum_{\beta=\{L,R\}}\sum_{\bq}\left [\omega_{\bq} b^{\dagger}_{\beta \bq}b_{\beta \bq}+\nonumber \right .\\
&+&\left  . \lambda_{\bq}(b^{\dagger}_{\beta \bq}+b_{\beta \bq})\,(a+a^{\dagger}) \right ].
\end{eqnarray}
Here $b^{\dagger}_{\beta \bq}$ is the creation creation operator of a photonic mode $\bq$ in the transmission line $\beta=\{L,R\}$ with photon energy $\omega_\bq$.
Inside the resonator the dissipation of energy occurs through two channels: 

({\it i}) capacitive coupling to the external photon bath 
with strength $\lambda_{\bq}$. In this channel the photons leak out of the cavity at a rate \cite{Schiro.14} 
\begin{eqnarray}
\cJ(\w)&=&\pi \sum_{\bq}\lambda_\bq^2 \delta(\omega-\omega_\bq)\nonumber\\
& \simeq & 2\pi\,\alpha\,\w\, e^{-\w/\w_c}\, ,
\label{eq:diss}
\end{eqnarray}
with $\w_c$ some energy cutoff.  In most of the cases, as the driving frequency
is close to the resonant frequency of the cavity, $\omega\simeq \omega_{0}$, the 
escape rate of the cavity $ \kappa$ is approximately constant and given by $\kappa =2 \pi\, \alpha\, \omega_0$, where $\alpha$ is a dimensionless constant.  

({\it ii}) The electron-photon interaction is the second channel of 
decay~\cite{Dmytruk.15a, Frey.12}. Besides the dissipative effects~\cite{Deng.15b},
it can also cause a shift of the resonance frequency. It was observed  that 
the quantum shot noise of the coherent conductor under the ac-bias can squeeze 
the photonic field~\cite{Gasse.13}. Recent theoretical 
studies~\cite{Dmytruk.15a} indicate that in the limit when the 
charge susceptibility is small, i.e. $|\Pi(\omega)|\ll \kappa$, the ratio $\Pi'(\omega)/ \kappa$
aproximates the phase shift and $\Pi''(\omega)/ \kappa$ corresponds to the cavity peak 
broadening, where the primed and double-primed quantities are the real and imaginary parts of the charge susceptibility.

\subsection{Green's function for the photons}\label{sec:Photons}

In this section we present the results for the spectral function $A_{a+a^\dagger}(\w)$ 
of the photonic operator $a+a^\dagger$. This quantity is needed for the evaluation of the microwave transmission across the cavity. 
The retarded Green's function for the photons is defined as
\be
\cD^R(t)= -i\Theta(t)\left \langle [a(t)+a^\dagger(t), a(0)+a^\dagger(0) ]\right \rangle.
\label{eq:DR}
\ee
The spectral function is obtained in terms of its Fourier transform $\cD^R(\w)$ as
\be
A_{a+a^\dagger}(\w) = -{1\over \pi} {\rm Im}\, \cD^R(\w).
\ee
In the absence of any sources of dissipation, the non-interacting correlator is
\be
\cD^R_0(\w)=\frac{2\w_0}{\w^2-\w_0^2+i 0^+}.
\ee
There are several ways to compute the spectral function, either directly from the NRG,
or by including the self-energy correction, as was originally proposed  for the Anderson model in Ref.~\cite{Bulla.98}, 
and later applied to the Anderson-Holstein model~\cite{Jeon.03}. 
We have used the latter approach, as the sum rule is better satisfied. 
The method consists in using the equation of motion for some 
correlators in order to  evaluate  the 
photonic self-energy (see Appendix~\ref{app:sf} for details), and then the use of Dyson equation to compute the full Green's function
\be
\cD^R(\w)^{-1} = \cD^R_0(\w)^{-1}-\Sigma_{\rm el}(\w)-\Sigma_{\rm ph} (\w) \; .\label{eq:Dyson}
\ee
Here, $\Sigma_{\rm ph} (\w)$ incorporates dissipation effects from the photon system only, coming from the coupling
of the cavity to the external modes, 
$\Sigma_{\rm ph} (\w) = -i\,\kappa/2$.
 \begin{figure}[!tb]
\includegraphics[width=0.9\columnwidth]{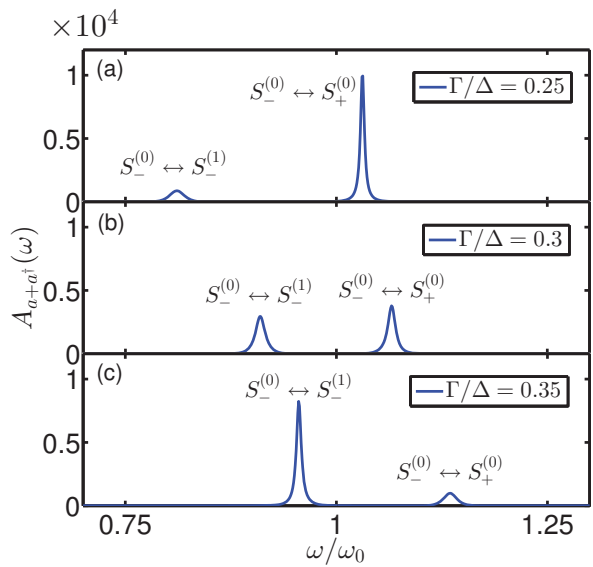}
\caption{(Color online) Spectral function $A_{a+a^\dagger}(\w)$ in different regimes. 
Each peak is associated with a given transition as indicated. 
 The other parameters 
correspond to those in Fig.~\ref{fig:levels}. We have used $\alpha=5\times 10^{-4}$ in 
Eq.~\eqref{eq:diss}.}
\label{fig:A_aa}
\end{figure}
In Eq.~\eqref{eq:Dyson}, the part of the self-energy coming from the light-matter interaction is obtained  as
\be
\Sigma_{\rm el}(\w) = g\frac{\cF^R(\w)}{\cD^R(\w)}\;,
\ee
where $\cF(\w)$ is the Fourier transform of the bosonic Green's function: 
$\cF^R(t)= -i\,\Theta(t)\left \langle [\hn(t), a(0)+a^\dagger(0) ]\right \rangle $.

Notice that if the ground state of the electronic system is the doublet, 
the photons remain decoupled, and implicitly $\Sigma_{\rm el}$ vanishes. In this 
case, the transmission across the cavity is not affected by  the electronic system. In this situation, no matter how close the avoided level crossing is to the QCP, there will always be just a single peak in the spectral density. Therefore, the discussion that follows refers mostly to the singlet side of the QCP. 
In our approach we have computed $\Sigma_{\rm el} 
(\w)$ exactly, 
by evaluating $\cF^R(\w)$ and a first approximation for $\cD^R(\w)$ at the level of NRG, and by using a constant value for the photonic part.

In Fig.~\ref{fig:A_aa} we present the NRG results for $A_{a+a^\dagger}(\w)$, when 
the system has a singlet ground state.
When the system 
is off-resonance, there is always one peak in $A_{a+a^{\dagger}}(\w)$ formed at 
$\omega\sim \omega_{0}$, which corresponds to the  transition from the ground 
state to the first photonic level with a width  $\kappa$ (panels (a) and (c) in 
Fig.~\ref{fig:A_aa}). This peak is mainly determined by $\cD_0^R(\w)$, and gives the spectral density of the non-interacting cavity. 
The other peak, much smaller in amplitude and shifted away from $\omega_0$, is due to 
the electron-photon interaction. Its position is given by the energy 
of the singlet Shiba state within the gap and its height is proportional to the 
weight of the photonic transition amplitude between the ground and the
excited state.
In Fig.~\ref{fig:A_aa}(b) we represent the 
situation when the system is at resonance: $E_{S_{+}}\simeq \hbar\, \omega_{0}$.
Now, both transitions, corresponding to $ S_-^{(0)}\to S_{-}^{(1)}$ and $S_-^{(0)}\to S_
+^{(0)}$, are similar in amplitude and considerably shifted to either side of $\omega_{0}$.

\subsection{Transmission across the cavity}

In our setup we consider that a microwave pulse comes from the left side and is either transmitted across the cavity to the right side, 
or reflected back. This can be viewed  as a scattering problem and the response can be addressed in terms  
of the many body $S$-matrix $\hS $. 
Energy conservation implies that the transition amplitude between some incoming $\ket i $ and outgoing $\ket f$ scattering states
is related to the on-shell $T$-matrix through 
\be
\bra f \hS\ket i=\delta_{fi}+2\pi i \delta(E_f-E_i)\bra f\cT\ket i, 
\ee
where $\bra f\cT\ket i$ is the on-shell $T$-matrix. By applying the path integral formalism 
(see Appendix~\ref{app:T} for details)
we can relate the on-shell $T$-matrix to the Green's functions for the photons in the cavity
\begin{eqnarray}
{\rm Im} \cT(\w) &=&  \pi \lambda_{\bq_f} \lambda_{\bq_i}\, A_{a+a^\dagger}(\w),\nonumber\\
{\rm Re} \cT(\w) &=&   \lambda_{\bq_f} \lambda_{\bq_i}\,\int d\w' \frac{ A_{a+a^\dagger}(\w')}{\w -\w'}\;.
\label{eq:Tmatrix}
\end{eqnarray}
\begin{figure}[!tb]
\includegraphics[width=0.47\columnwidth]{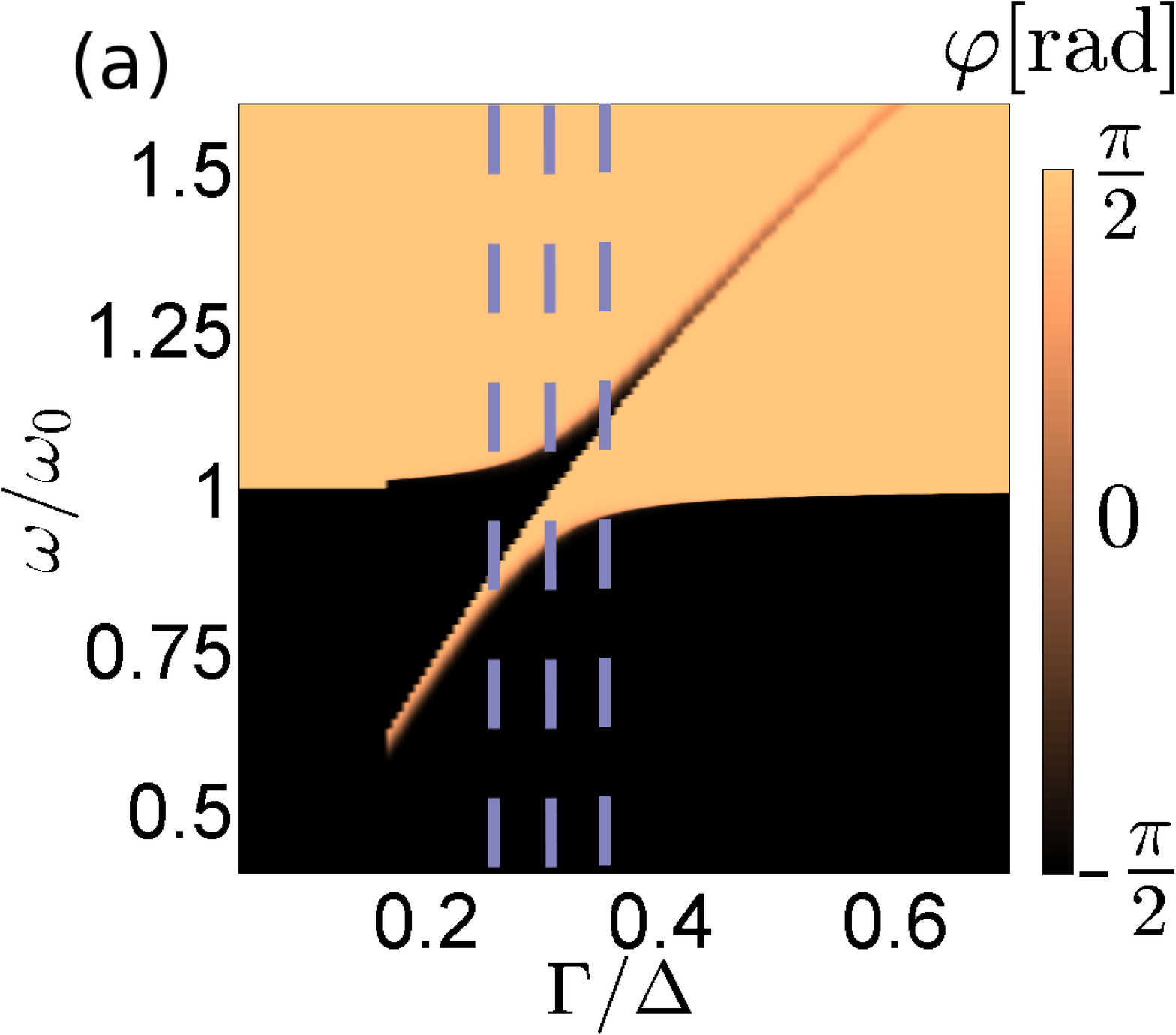}
\includegraphics[width=0.49\columnwidth]{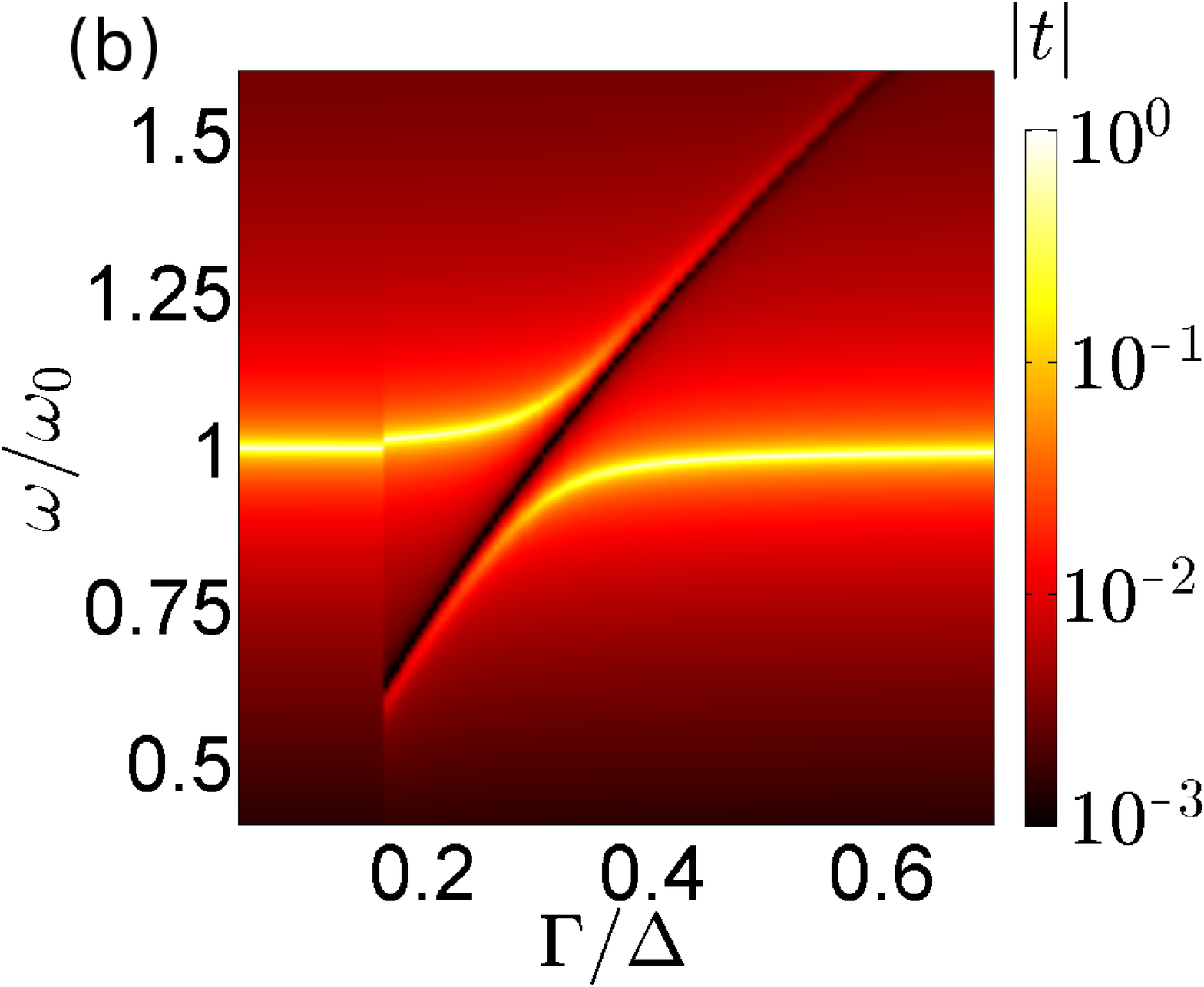}
\caption{(Color online) Density plots for the phase and absolute value of the transmission through the cavity. The three dashed lines correspond to the cuts along which the phase
and amplitude of the transmission are displayed in Figs.~\ref{fig:1D_phase} and \ref{fig:t_amp}.} 
\label{fig:2D_t}
\end{figure}

 \begin{figure}[tb]
\includegraphics[width=0.9\columnwidth]{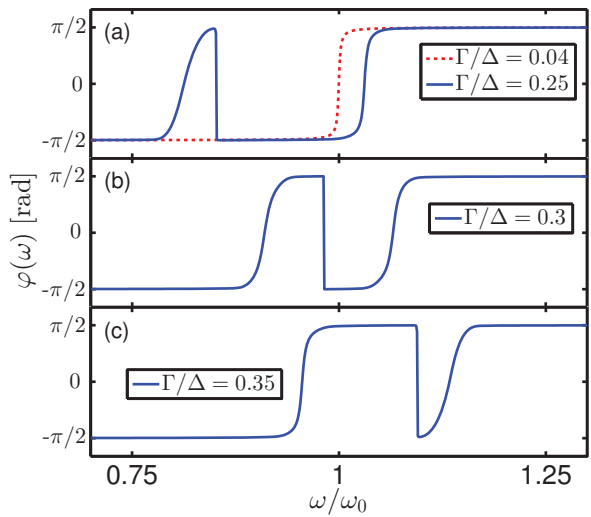}
\caption{(Color online) Phase of the transmission through the cavity 
when the system is in the singlet ground state, corresponding to $x>x_{\rm QCP}$.  
The three panels correspond to the vertical cuts in Fig.~\ref{fig:2D_t} (a). 
For $x<x_{\rm QCP}$ the system simply shows a jump of $\pi$ at $\omega/\omega_{0}=1$ 
(diplayed as a red dashed line in panel (a)).
} 
\label{fig:1D_phase}
\end{figure}

 \begin{figure}[!tb]
\includegraphics[width=0.9\columnwidth]{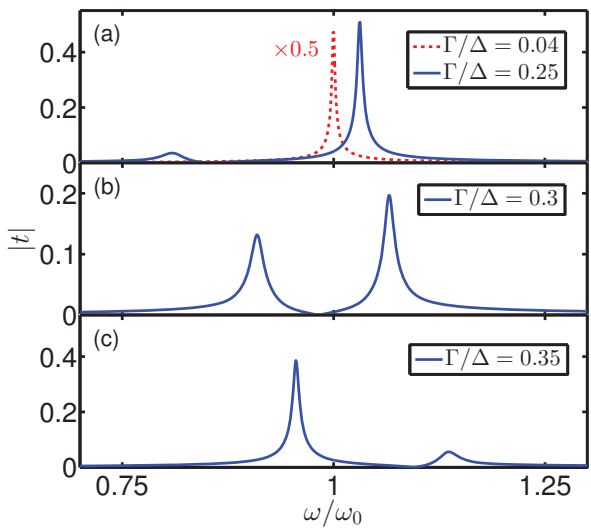}
\caption{(Color online) Absolute value for the amplitude of the transmission through the cavity for the same values of $\Gamma/\Delta$ as in Fig.~\ref{fig:1D_phase}. 
The dashes red line in panel (a) represent the transmission amplitude when the system is 
in a doublet ground state, and corresponds to the otherwise free cavity. 
}
\label{fig:t_amp}
\end{figure}

If we further assume that the density of states $\rho(\w)$ for the modes in the external transmission lines varies slowly with the frequency around $\omega=\omega_\bq$, and that the coupling $\lambda_\bq$
is approximately constant, we can write  the transmission
\begin{eqnarray}
\label{eq:tra}
t(\w)= 2\pi i\,\rho(\w) \cT(\w) = i\cJ(\w) \cD^R(\w)\label{eq:T_matrix}\;,
\end{eqnarray}
with $\cD^R(\w)$ the Fourier transform of $\cD^R(t)$ defined in Eq.~\eqref{eq:DR}.
We define the amplitude $ |t(\w)|$ and the phase 
$\varphi(\w)$ of the transmission appearing in Eq.~\eqref{eq:T_matrix} through
\be
t(\w) = |t(\w)|\, e^{i\varphi(\w)}\, .
\ee
In Fig.~\ref{fig:2D_t} we display a density plot of these two quantities, while 
in Figs.~\ref{fig:1D_phase} and \ref{fig:t_amp} we display the same quantities along 
three cuts at fixed $\Gamma/\Delta$ ratios. 

It is straightforward to understand the amplitude behavior, since $|t(\w)|$  follows closely the behavior of the spectral function $A_{a+a^{\dagger}}(\w)$,  except for some small asymmetric 
contributions coming from  the real part of the Green's function, which in the present calculations 
was computed by a Hilbert transform. The peaks in $|t(\w)|$ can be understood along the same lines as those for $A_{a+a^{\dagger}}(\w)$, which we discussed in Sec.~\ref{sec:Photons}. The important point we want to make here is that 
the Shiba states leave a clear fingerprint in this measurable quantity. 
By tuning 
the system parameters, the coupling can be resonant, and the strength of the interaction 
can be measured experimentally~\cite{Janvier.15}. If the avoided level crossing is tuned
close to the QCP in the singlet region, then, by crossing through the QCP into 
the doublet region (by tuning $\Gamma$ for example), the photonic transition between the ground and the excited states becomes blocked, and  
the signal coming from these states must disappear. In the doublet region, the 
transmission is represented by the red dashed line in Fig.~\ref{fig:t_amp}(a).
This offers us a unique approach to exactly pinpoint the QPT. 

The other quantity of interest that is accessible experimentally is the 
change in the phase of the transmission. In the absence of the electronic system, 
$\varphi(\w)$ shows a change of $\pi$ radians as a function of the incoming microwave  frequency $\w$ at the resonance frequency $\omega=\omega_{0}$. 
This is clearly visible in Fig.~\ref{fig:2D_t}(a) when $x<x_{\rm QCP}$, and the ground 
state is the doublet. In the doublet region, the phase will always present a single 
jump of $\pi$ (indicated by the dashed line in Fig.~\ref{fig:1D_phase}(a)). In the singlet region close to the avoided level crossing, the coupling to 
the electronic system becomes important and the 
self-energy $\Sigma_{\rm el}$ increases. Furthermore, the 
light-matter interaction introduces more phase
jumps of $\pm \pi$ each time an excited Shiba state (within the singlet channel) is crossed. 
In contrast to the absolute value of the transmission amplitude, for which the signal can be weak, the phase change could give a much clearer signature of the presence of the Shiba states. For a true bound state with zero width, the phase will always show an abrupt jump of $\pm \pi$, 
the smooth steps visible in Fig.~\ref{fig:1D_phase} being numerical artifacts 
due to the finite broadening of the bound states. 

%

\section{Concluding Remarks}

We have studied a hybrid system that consists of a quantum dot coupled to 
superconducting leads and also to a microwave cavity. In the absence of photons,
such an electronic system develops 
multiple Shiba states inside the superconducting gap~\cite{Shiba.68}. They have been recently investigated by using subgap 
spectroscopy~\cite{Gaass.14}.
The competition between them drives the system through a quantum phase transition. 
On one side of the transition, the ground state is a doublet, while on the other side 
it is a singlet. 

It is  accepted by now that coupling a mesoscopic device,
such as a tunneling junction or a quantum dot, to a cavity, 
provides a non-invasive way to explore electronic processes~\cite{Frey.12, Delbecq.11, Wallraff.04}. Along this line, in the present work we demonstrate that coupling a mesoscopic system, 
that supports Shiba states, to a microwave cavity, offers a
controllable way to monitor these localized states~\cite{Janvier.15} and their evolution across the 
quantum phase transition. 
If the ground state is a doublet, the cavity is insensitive to the presence of the 
electronic system, and the transmission is unaltered. If the system is tuned across 
the quantum critical point in the region where the ground state is a singlet, 
the coupling between the electronic system and the cavity leaves distinctive features
in the transmitted signal. 
The computational methodology was based on the numerical renomalization group approach 
supplemented by analytical calculations.

\begin{acknowledgments}
We are thankful to Gergely Zarand, Andras Palyi and Vidar Gudmundsson for insightful comments and lively discussions. This work was supported by the Romanian National Authority for Scientific Research and
Innovation, UEFISCDI, project number PN-II-RU-TE-2014-4-0432, and by the Hungarian research fund OTKA under grant No. K105149. 
\end{acknowledgments}
\appendix

\section{Numerical renormalization group approach}\label{app:NRG}
We solve the Hamiltonian in Eq. \eqref{eq:H_sys} by using the numerical renormalization group.
We shall discuss the situation with the left-right symmetry and $\varphi_{\alpha}=0$, 
with $\alpha = \{L,R\}$ 
(no phase difference between the superconducting gaps on the left/right leads) . 
We  first perform a unitary transformation \cite{Hecht.08} 
and introduce a new basis
\begin{eqnarray}
\left(
\begin{array}{cc}
c_{e\bf k\sigma} \\
c_{o\bf k\sigma}
\end{array}
\right)&=&\frac{1}{\sqrt{2}}
\left(
\begin{array}{cc}
1 & 1 \\
-1  & 1
\end{array}
\right)
\left(
\begin{array}{cc}
c_{L\bf k\sigma}  \nonumber\\
c_{R\bf k\sigma}
\end{array}
\right)\; .
\end{eqnarray}
The advantage of this even-odd  basis over the left-right one 
is that the odd channel becomes decoupled and the 'impurity' remains coupled
only to the even channel. Performing a logarithmic discretization 
of the conduction band by
using a parameter $\Lambda>1$, 
 followed by a tridiagonalization procedure that maps
the problem on the Wilson chain~\cite{Wilson.75}, the leads Hamiltonian transforms to
\begin{eqnarray}
H_{\mathrm{leads}}& = & \sum_{n=0}^{\infty}\sum_{\sigma}\Big ( t_{n}\;
f^{\dagger}_{n \sigma}f_{n+1 \sigma}+ 
\Delta f^{\dagger}_{n\uparrow}f^{\dagger}_{n \downarrow}+\mathrm{H.c.}\Big )\, ,
\nonumber
\end{eqnarray}
where $t_{n}$ is the hopping between nearest neighbor sites along the Wilson chain, 
which decreases exponentially fast,
$t_{n}\sim \Lambda^{-n/2}$, and $f^{\dagger}_{n\sigma}$ is the fermionic creation 
operator at the n-th site. 
 The 'impurity' is coupled to the first site of the Wilson
chain, by the tunneling Hamiltonian
\begin{eqnarray}
H_{\mathrm{tun}}& = &\sum_{ \sigma}\left (\sqrt{2}V\;
f^{\dagger}_{0 \sigma} d_{\sigma} 
+\mathrm{H.c.} \right )\;.
\end{eqnarray}
The 'impurity' is a complex object that takes into account all 
the bosonic states for the photons. In our calculation we have considered only five such photonic states. Increasing this number does not visibly affect our results. 
The  Hamiltonian for the 'impurity'

\begin{equation}
H_{\rm impurity}= H_{\rm dot}+H_{\rm cav}+H_{\rm int}\, 
\end{equation}
remains unaffected  by the unitary transformation, and is the sum of Hamiltonian terms discussed in Sec.~\ref{sec:Hamiltonian}
and given in Eqs.~\eqref{eq:H_dot},  \eqref{eq:H_int} and \eqref{eq:H_cav}.
The model is SU(2) invariant, as the total Hamiltonian commutes with the spin operator, 
which allows us to keep track of this symmetry when indexing the eigenstates of the Hamiltonian. 


\section{Photon Spectral Function}\label{app:sf}
In this appendix we shall give a brief derivation of Dyson Eq. \eqref{eq:Dyson}. The 
goal is to compute the bosonic Green's function from eq.~\eqref{eq:DR} accurately. 
We start by defining the retarded Green's function of two bosonic operators ${\cal O}_1$ and ${\cal O}_2$ as
\begin{eqnarray}\label{eq:D_d}
\cD_{{\cal O} _1,{\cal O} _2}^R(t) &=&-i\theta(t)\langle \left[{\cal O}_1(t),{\cal O}_2(0) \right]\rangle\,.
\end{eqnarray}
Its Fourier transform, $\cD_{{\cal O} _1,{\cal O} _2}^R(\omega)$, satisfies the equation 
of motion
\begin{eqnarray}
\omega\;\cD_{{\cal O} _1,{\cal O} _2}^R(\omega)&=&\langle \left[{\cal O} _1,{\cal O} _2 \right] \rangle-\cD_{{\cal O} _1,\left[{\cal O} _2,H\right ]}^R(\omega).
\end{eqnarray}
 Applying this formula successively, first for ${\cal O}_1=a+a^{\dagger}$, ${\cal O}_2=a+a^{\dagger}$, and then for ${\cal O}_1=a^{\dagger}-a$, ${\cal O}_2=a+a^{\dagger}$, we 
 immediately obtain eq.~\eqref{eq:Dyson} \cite{Jeon.03}.
Notice that the superconducting gap does not enter explicitly in Eq. \eqref{eq:Dyson}. The correlator $\cF^{R}(\w)$ is computed directly from the NRG. 
\section{T-matrix for the external photon bath}
\label{app:T}
In this appendix we derive the expression for the $T$-matrix for the external photon bath. It is most convenient
to work with a path integral formalism. Our goal is to express the full Green's function for the 
bosonic modes of the photons in the external transmission lines. In this approach, the electronic part plays no role, and 
can be integrated out from the beginning. The partition and the generating functions can be constructed by 
replacing the bosonic fields by time dependent complex variables~\cite{Negele}:
$\{b_{\beta q}, b^{\dagger}_{\beta q}, a, a^{\dagger} \}\to \{b_{\beta q}, \bar b_{\beta q}, a, \bar a \}$. 
For the generating function it follows:
\begin{equation}
{\cal Z}[\eta, \bar \eta] = \int D[b\,\bar b]\; D[a\,\bar a]\, e^{-i\, {\cal S}[b,\bar b, a, \bar a]+i\, \bar \eta\cdot b +i\,\bar b\cdot\eta }\; ,
\label{eq:Z}
\end{equation}
with the total action consisting of several terms: $\cS =\cS_{\rm bath}+\cS_{\rm cav}+\cS_{\rm int}$. 
Here we have used the notation: 
$\bar\eta\cdot  b = \sum_{{\beta, q}} \int dt \bar\eta_{\beta q}(t)b_{\beta q}(t)$. The 
non-interacting part \c{$\cS_{\rm bath}$} describes the photons in the external bath and is given by
\begin{equation}
\cS_{\rm bath} = -\sum_{{\beta=\{ L,R\}}}\sum_{q}\int dt\, dt'\, \bar b_{\beta q}(t)\,{\cal D}_{0,\beta q}^{-1}(t-t')\, b_{\beta q}(t'),
\nonumber
\end{equation}
with ${\cal D}_{0,\beta q}(t-t')$ the corresponding non-interacting Green's function. The cavity is  described by the action
\begin{equation}
\cS_{\rm cav} = -\int dt\, dt' \bar a(t)\,{\cal D}^{-1}(t-t')\,  a(t').
\nonumber
\end{equation}
Here  ${\cal D}(t-t')$ is the Green's function for the photons in the cavity, that contains 
the self-energy contribution by integrating the electronic part. 
The interaction part describes the hybridization of the modes in the bath with those in the cavity
\begin{equation}
\cS_{\rm int} = \sum_{\beta=\{L,R \}}\sum_{q}\lambda_{q}\int dt 
\Big [ \bar a(t)\, b_{\beta q}(t)+ c.c\, \Big ].
\nonumber
\end{equation}
The full Green's function for the external modes can be written as a functional differentiation
\begin{equation}
\cD_{\beta q, \beta' q'} = -i \frac{\delta^{2} \ln \cZ[\eta, \bar \eta]}{\delta \bar\eta_{\beta q}(t)\,\delta\eta_{\beta' q'}(t')}\Big |_{\eta,\bar \eta\to 0}.
\label{eq:diff}
\end{equation}
Then the expression for the $T$-matrix can be obtained by shifting the integration argument 
in the exponent of the generating function in \eqref{eq:Z}
\begin{equation}
b_{\beta q}\to b_{\beta q}-\int dt' \cD_{0, \beta q} (t-t') \eta_{\beta q}(t'),
\nonumber
\end{equation}
followed by the differentiation~\eqref{eq:diff}. Finally, the full Green's function is:
\begin{multline}
\cD_{\beta q_{f}, \beta' q_{i}}(t-t')= \delta_{\beta\beta'} \delta_{q_{f} q_{i}} \cD_{0,\beta q_{i}}(t-t')-\\
- i\, \lambda_{q_{f}}\lambda_{q_{i}}\int dt_{1}\,dt_{1}'\, \cD_{0,\beta q_{f}}(t-t_{1})\, \cD(t_{1}-t_{1}')\,\cD_{0,\beta' q_{i}}(t_{1}'-t')\; ,
\nonumber
\end{multline}
which allows us to identify the expression for the $T$-matrix, given in Eq.~\eqref{eq:Tmatrix}.
\bibliography{references}

\begin{thebibliography}{38}%
\makeatletter
\providecommand \@ifxundefined [1]{%
 \@ifx{#1\undefined}
}%
\providecommand \@ifnum [1]{%
 \ifnum #1\expandafter \@firstoftwo
 \else \expandafter \@secondoftwo
 \fi
}%
\providecommand \@ifx [1]{%
 \ifx #1\expandafter \@firstoftwo
 \else \expandafter \@secondoftwo
 \fi
}%
\providecommand \natexlab [1]{#1}%
\providecommand \enquote  [1]{``#1''}%
\providecommand \bibnamefont  [1]{#1}%
\providecommand \bibfnamefont [1]{#1}%
\providecommand \citenamefont [1]{#1}%
\providecommand \href@noop [0]{\@secondoftwo}%
\providecommand \href [0]{\begingroup \@sanitize@url \@href}%
\providecommand \@href[1]{\@@startlink{#1}\@@href}%
\providecommand \@@href[1]{\endgroup#1\@@endlink}%
\providecommand \@sanitize@url [0]{\catcode `\\12\catcode `\$12\catcode
  `\&12\catcode `\#12\catcode `\^12\catcode `\_12\catcode `\%12\relax}%
\providecommand \@@startlink[1]{}%
\providecommand \@@endlink[0]{}%
\providecommand \url  [0]{\begingroup\@sanitize@url \@url }%
\providecommand \@url [1]{\endgroup\@href {#1}{\urlprefix }}%
\providecommand \urlprefix  [0]{URL }%
\providecommand \Eprint [0]{\href }%
\providecommand \doibase [0]{http://dx.doi.org/}%
\providecommand \selectlanguage [0]{\@gobble}%
\providecommand \bibinfo  [0]{\@secondoftwo}%
\providecommand \bibfield  [0]{\@secondoftwo}%
\providecommand \translation [1]{[#1]}%
\providecommand \BibitemOpen [0]{}%
\providecommand \bibitemStop [0]{}%
\providecommand \bibitemNoStop [0]{.\EOS\space}%
\providecommand \EOS [0]{\spacefactor3000\relax}%
\providecommand \BibitemShut  [1]{\csname bibitem#1\endcsname}%
\let\auto@bib@innerbib\@empty
\bibitem [{\citenamefont {Xiang}\ \emph {et~al.}(2013)\citenamefont {Xiang},
  \citenamefont {Ashhab}, \citenamefont {You},\ and\ \citenamefont
  {Nori}}]{Xiang.13}%
  \BibitemOpen
  \bibfield  {author} {\bibinfo {author} {\bibfnamefont {Z.-L.}\ \bibnamefont
  {Xiang}}, \bibinfo {author} {\bibfnamefont {S.}~\bibnamefont {Ashhab}},
  \bibinfo {author} {\bibfnamefont {J.~Q.}\ \bibnamefont {You}}, \ and\
  \bibinfo {author} {\bibfnamefont {F.}~\bibnamefont {Nori}},\ }\href@noop {}
  {\bibfield  {journal} {\bibinfo  {journal} {Rev. Mod. Phys.}\ }\textbf
  {\bibinfo {volume} {85}},\ \bibinfo {pages} {623} (\bibinfo {year}
  {2013})}\BibitemShut {NoStop}%
\bibitem [{\citenamefont {Wallraff}\ \emph {et~al.}(2004)\citenamefont
  {Wallraff}, \citenamefont {Schuster}, \citenamefont {Blais}, \citenamefont
  {Frunzio}, \citenamefont {Huang}, \citenamefont {Majer}, \citenamefont
  {Kumar}, \citenamefont {Girvin},\ and\ \citenamefont
  {Schoelkopf}}]{Wallraff.04}%
  \BibitemOpen
  \bibfield  {author} {\bibinfo {author} {\bibfnamefont {A.}~\bibnamefont
  {Wallraff}}, \bibinfo {author} {\bibfnamefont {D.~I.}\ \bibnamefont
  {Schuster}}, \bibinfo {author} {\bibfnamefont {A.}~\bibnamefont {Blais}},
  \bibinfo {author} {\bibfnamefont {L.}~\bibnamefont {Frunzio}}, \bibinfo
  {author} {\bibfnamefont {R.~S.}\ \bibnamefont {Huang}}, \bibinfo {author}
  {\bibfnamefont {J.}~\bibnamefont {Majer}}, \bibinfo {author} {\bibfnamefont
  {S.}~\bibnamefont {Kumar}}, \bibinfo {author} {\bibfnamefont {S.~M.}\
  \bibnamefont {Girvin}}, \ and\ \bibinfo {author} {\bibfnamefont {R.~J.}\
  \bibnamefont {Schoelkopf}},\ }\href@noop {} {\bibfield  {journal} {\bibinfo
  {journal} {Nature}\ }\textbf {\bibinfo {volume} {431}},\ \bibinfo {pages}
  {162} (\bibinfo {year} {2004})}\BibitemShut {NoStop}%
\bibitem [{\citenamefont {Frey}\ \emph {et~al.}(2011)\citenamefont {Frey},
  \citenamefont {Leek}, \citenamefont {Beck}, \citenamefont {Ensslin},
  \citenamefont {Wallraff},\ and\ \citenamefont {Ihn}}]{Frey.11}%
  \BibitemOpen
  \bibfield  {author} {\bibinfo {author} {\bibfnamefont {T.}~\bibnamefont
  {Frey}}, \bibinfo {author} {\bibfnamefont {P.~J.}\ \bibnamefont {Leek}},
  \bibinfo {author} {\bibfnamefont {M.}~\bibnamefont {Beck}}, \bibinfo {author}
  {\bibfnamefont {K.}~\bibnamefont {Ensslin}}, \bibinfo {author} {\bibfnamefont
  {A.}~\bibnamefont {Wallraff}}, \ and\ \bibinfo {author} {\bibfnamefont
  {T.}~\bibnamefont {Ihn}},\ }\href@noop {} {\bibfield  {journal} {\bibinfo
  {journal} {Appl. Phys. Lett.}\ }\textbf {\bibinfo {volume} {98}},\ \bibinfo
  {pages} {262105} (\bibinfo {year} {2011})}\BibitemShut {NoStop}%
\bibitem [{\citenamefont {Rundquist}\ \emph {et~al.}(2011)\citenamefont
  {Rundquist}, \citenamefont {Majumdar},\ and\ \citenamefont
  {Vuckovic}}]{Rundquist.11}%
  \BibitemOpen
  \bibfield  {author} {\bibinfo {author} {\bibfnamefont {A.}~\bibnamefont
  {Rundquist}}, \bibinfo {author} {\bibfnamefont {A.}~\bibnamefont {Majumdar}},
  \ and\ \bibinfo {author} {\bibfnamefont {J.}~\bibnamefont {Vuckovic}},\
  }\href@noop {} {\bibfield  {journal} {\bibinfo  {journal} {Applied Physics
  Letters}\ }\textbf {\bibinfo {volume} {99}} (\bibinfo {year}
  {2011})}\BibitemShut {NoStop}%
\bibitem [{\citenamefont {Delbecq}\ \emph {et~al.}(2011)\citenamefont
  {Delbecq}, \citenamefont {Schmitt}, \citenamefont {Parmentier}, \citenamefont
  {Roch}, \citenamefont {Viennot}, \citenamefont {F\`eve}, \citenamefont
  {Huard}, \citenamefont {Mora}, \citenamefont {Cottet},\ and\ \citenamefont
  {Kontos}}]{Delbecq.11}%
  \BibitemOpen
  \bibfield  {author} {\bibinfo {author} {\bibfnamefont {M.~R.}\ \bibnamefont
  {Delbecq}}, \bibinfo {author} {\bibfnamefont {V.}~\bibnamefont {Schmitt}},
  \bibinfo {author} {\bibfnamefont {F.~D.}\ \bibnamefont {Parmentier}},
  \bibinfo {author} {\bibfnamefont {N.}~\bibnamefont {Roch}}, \bibinfo {author}
  {\bibfnamefont {J.~J.}\ \bibnamefont {Viennot}}, \bibinfo {author}
  {\bibfnamefont {G.}~\bibnamefont {F\`eve}}, \bibinfo {author} {\bibfnamefont
  {B.}~\bibnamefont {Huard}}, \bibinfo {author} {\bibfnamefont
  {C.}~\bibnamefont {Mora}}, \bibinfo {author} {\bibfnamefont {A.}~\bibnamefont
  {Cottet}}, \ and\ \bibinfo {author} {\bibfnamefont {T.}~\bibnamefont
  {Kontos}},\ }\href@noop {} {\bibfield  {journal} {\bibinfo  {journal} {Phys.
  Rev. Lett.}\ }\textbf {\bibinfo {volume} {107}},\ \bibinfo {pages} {256804}
  (\bibinfo {year} {2011})}\BibitemShut {NoStop}%
\bibitem [{\citenamefont {Deng}\ \emph
  {et~al.}(2015{\natexlab{a}})\citenamefont {Deng}, \citenamefont {Henriet},
  \citenamefont {Li}, \citenamefont {Li}, \citenamefont {Cao}, \citenamefont
  {Xiao}, \citenamefont {Guo}, \citenamefont {Schiro}, \citenamefont {Le~Hur},\
  and\ \citenamefont {Guo}}]{Deng.15}%
  \BibitemOpen
  \bibfield  {author} {\bibinfo {author} {\bibfnamefont {G.-W.}\ \bibnamefont
  {Deng}}, \bibinfo {author} {\bibfnamefont {D.~W.}\ \bibnamefont {Henriet}},
  \bibinfo {author} {\bibfnamefont {S.-X.}\ \bibnamefont {Li}}, \bibinfo
  {author} {\bibfnamefont {H.-O.}\ \bibnamefont {Li}}, \bibinfo {author}
  {\bibfnamefont {G.}~\bibnamefont {Cao}}, \bibinfo {author} {\bibfnamefont
  {M.}~\bibnamefont {Xiao}}, \bibinfo {author} {\bibfnamefont {G.-C.}\
  \bibnamefont {Guo}}, \bibinfo {author} {\bibfnamefont {M.}~\bibnamefont
  {Schiro}}, \bibinfo {author} {\bibfnamefont {K.}~\bibnamefont {Le~Hur}}, \
  and\ \bibinfo {author} {\bibfnamefont {G.-P.}\ \bibnamefont {Guo}},\
  }\href@noop {} {\bibfield  {journal} {\bibinfo  {journal}
  {arXiv:1509.06141v2}\ } (\bibinfo {year} {2015}{\natexlab{a}})}\BibitemShut
  {NoStop}%
\bibitem [{\citenamefont {Janvier}\ \emph {et~al.}(2015)\citenamefont
  {Janvier}, \citenamefont {Tosi}, \citenamefont {Bretheau}, \citenamefont
  {Girit}, \citenamefont {Stern}, \citenamefont {Bertet}, \citenamefont
  {Joyez}, \citenamefont {Vion}, \citenamefont {Esteve}, \citenamefont
  {Goffman}, \citenamefont {Pothier},\ and\ \citenamefont
  {Urbina}}]{Janvier.15}%
  \BibitemOpen
  \bibfield  {author} {\bibinfo {author} {\bibfnamefont {C.}~\bibnamefont
  {Janvier}}, \bibinfo {author} {\bibfnamefont {L.}~\bibnamefont {Tosi}},
  \bibinfo {author} {\bibfnamefont {L.}~\bibnamefont {Bretheau}}, \bibinfo
  {author} {\bibfnamefont {C.~O.}\ \bibnamefont {Girit}}, \bibinfo {author}
  {\bibfnamefont {M.}~\bibnamefont {Stern}}, \bibinfo {author} {\bibfnamefont
  {P.}~\bibnamefont {Bertet}}, \bibinfo {author} {\bibfnamefont
  {P.}~\bibnamefont {Joyez}}, \bibinfo {author} {\bibfnamefont
  {D.}~\bibnamefont {Vion}}, \bibinfo {author} {\bibfnamefont {D.}~\bibnamefont
  {Esteve}}, \bibinfo {author} {\bibfnamefont {M.~F.}\ \bibnamefont {Goffman}},
  \bibinfo {author} {\bibfnamefont {H.}~\bibnamefont {Pothier}}, \ and\
  \bibinfo {author} {\bibfnamefont {C.}~\bibnamefont {Urbina}},\ }\href@noop {}
  {\bibfield  {journal} {\bibinfo  {journal} {Science}\ }\textbf {\bibinfo
  {volume} {349}},\ \bibinfo {pages} {1199} (\bibinfo {year}
  {2015})}\BibitemShut {NoStop}%
\bibitem [{\citenamefont {Frey}\ \emph {et~al.}(2012)\citenamefont {Frey},
  \citenamefont {Leek}, \citenamefont {Beck}, \citenamefont {Blais},
  \citenamefont {Ihn}, \citenamefont {Ensslin},\ and\ \citenamefont
  {Wallraff}}]{Frey.12}%
  \BibitemOpen
  \bibfield  {author} {\bibinfo {author} {\bibfnamefont {T.}~\bibnamefont
  {Frey}}, \bibinfo {author} {\bibfnamefont {P.~J.}\ \bibnamefont {Leek}},
  \bibinfo {author} {\bibfnamefont {M.}~\bibnamefont {Beck}}, \bibinfo {author}
  {\bibfnamefont {A.}~\bibnamefont {Blais}}, \bibinfo {author} {\bibfnamefont
  {T.}~\bibnamefont {Ihn}}, \bibinfo {author} {\bibfnamefont {K.}~\bibnamefont
  {Ensslin}}, \ and\ \bibinfo {author} {\bibfnamefont {A.}~\bibnamefont
  {Wallraff}},\ }\href@noop {} {\bibfield  {journal} {\bibinfo  {journal}
  {Phys. Rev. Lett.}\ }\textbf {\bibinfo {volume} {108}},\ \bibinfo {pages}
  {046807} (\bibinfo {year} {2012})}\BibitemShut {NoStop}%
\bibitem [{\citenamefont {van Loo}\ \emph {et~al.}(2013)\citenamefont {van
  Loo}, \citenamefont {Fedorov}, \citenamefont {Lalumière}, \citenamefont
  {Sanders}, \citenamefont {Blais},\ and\ \citenamefont
  {Wallraff}}]{vanLoo.13}%
  \BibitemOpen
  \bibfield  {author} {\bibinfo {author} {\bibfnamefont {A.~F.}\ \bibnamefont
  {van Loo}}, \bibinfo {author} {\bibfnamefont {A.}~\bibnamefont {Fedorov}},
  \bibinfo {author} {\bibfnamefont {K.}~\bibnamefont {Lalumière}}, \bibinfo
  {author} {\bibfnamefont {B.~C.}\ \bibnamefont {Sanders}}, \bibinfo {author}
  {\bibfnamefont {A.}~\bibnamefont {Blais}}, \ and\ \bibinfo {author}
  {\bibfnamefont {A.}~\bibnamefont {Wallraff}},\ }\href@noop {} {\bibfield
  {journal} {\bibinfo  {journal} {Science}\ }\textbf {\bibinfo {volume}
  {342}},\ \bibinfo {pages} {1494} (\bibinfo {year} {2013})}\BibitemShut
  {NoStop}%
\bibitem [{\citenamefont {Kulkarni}\ \emph {et~al.}(2014)\citenamefont
  {Kulkarni}, \citenamefont {Cotlet},\ and\ \citenamefont
  {T\"ureci}}]{Kulkarni.14}%
  \BibitemOpen
  \bibfield  {author} {\bibinfo {author} {\bibfnamefont {M.}~\bibnamefont
  {Kulkarni}}, \bibinfo {author} {\bibfnamefont {O.}~\bibnamefont {Cotlet}}, \
  and\ \bibinfo {author} {\bibfnamefont {H.~E.}\ \bibnamefont {T\"ureci}},\
  }\href@noop {} {\bibfield  {journal} {\bibinfo  {journal} {Phys. Rev. B}\
  }\textbf {\bibinfo {volume} {90}},\ \bibinfo {pages} {125402} (\bibinfo
  {year} {2014})}\BibitemShut {NoStop}%
\bibitem [{\citenamefont {Blais}\ \emph {et~al.}(2004)\citenamefont {Blais},
  \citenamefont {Huang}, \citenamefont {Wallraff}, \citenamefont {Girvin},\
  and\ \citenamefont {Schoelkopf}}]{Blais.04}%
  \BibitemOpen
  \bibfield  {author} {\bibinfo {author} {\bibfnamefont {A.}~\bibnamefont
  {Blais}}, \bibinfo {author} {\bibfnamefont {R.-S.}\ \bibnamefont {Huang}},
  \bibinfo {author} {\bibfnamefont {A.}~\bibnamefont {Wallraff}}, \bibinfo
  {author} {\bibfnamefont {S.~M.}\ \bibnamefont {Girvin}}, \ and\ \bibinfo
  {author} {\bibfnamefont {R.~J.}\ \bibnamefont {Schoelkopf}},\ }\href@noop {}
  {\bibfield  {journal} {\bibinfo  {journal} {Phys. Rev. A}\ }\textbf {\bibinfo
  {volume} {69}},\ \bibinfo {pages} {062320} (\bibinfo {year}
  {2004})}\BibitemShut {NoStop}%
\bibitem [{\citenamefont {Cottet}\ \emph {et~al.}(2015)\citenamefont {Cottet},
  \citenamefont {Kontos},\ and\ \citenamefont {Dou\ifmmode~\mbox{\c{c}}\else
  \c{c}\fi{}ot}}]{Cottet.15}%
  \BibitemOpen
  \bibfield  {author} {\bibinfo {author} {\bibfnamefont {A.}~\bibnamefont
  {Cottet}}, \bibinfo {author} {\bibfnamefont {T.}~\bibnamefont {Kontos}}, \
  and\ \bibinfo {author} {\bibfnamefont {B.}~\bibnamefont
  {Dou\ifmmode~\mbox{\c{c}}\else \c{c}\fi{}ot}},\ }\href@noop {} {\bibfield
  {journal} {\bibinfo  {journal} {Phys. Rev. B}\ }\textbf {\bibinfo {volume}
  {91}},\ \bibinfo {pages} {205417} (\bibinfo {year} {2015})}\BibitemShut
  {NoStop}%
\bibitem [{\citenamefont {Sk\"oldberg}\ \emph {et~al.}(2008)\citenamefont
  {Sk\"oldberg}, \citenamefont {L\"ofwander}, \citenamefont {Shumeiko},\ and\
  \citenamefont {Fogelstr\"om}}]{Skoldberg.08}%
  \BibitemOpen
  \bibfield  {author} {\bibinfo {author} {\bibfnamefont {J.}~\bibnamefont
  {Sk\"oldberg}}, \bibinfo {author} {\bibfnamefont {T.}~\bibnamefont
  {L\"ofwander}}, \bibinfo {author} {\bibfnamefont {V.~S.}\ \bibnamefont
  {Shumeiko}}, \ and\ \bibinfo {author} {\bibfnamefont {M.}~\bibnamefont
  {Fogelstr\"om}},\ }\href@noop {} {\bibfield  {journal} {\bibinfo  {journal}
  {Phys. Rev. Lett.}\ }\textbf {\bibinfo {volume} {101}},\ \bibinfo {pages}
  {087002} (\bibinfo {year} {2008})}\BibitemShut {NoStop}%
\bibitem [{\citenamefont {Dmytruk}\ \emph
  {et~al.}(2015{\natexlab{a}})\citenamefont {Dmytruk}, \citenamefont {Trif},\
  and\ \citenamefont {Simon}}]{Dmytruk.15a}%
  \BibitemOpen
  \bibfield  {author} {\bibinfo {author} {\bibfnamefont {O.}~\bibnamefont
  {Dmytruk}}, \bibinfo {author} {\bibfnamefont {M.}~\bibnamefont {Trif}}, \
  and\ \bibinfo {author} {\bibfnamefont {P.}~\bibnamefont {Simon}},\
  }\href@noop {} {\bibfield  {journal} {\bibinfo  {journal} {Phys. Rev. B}\
  }\textbf {\bibinfo {volume} {92}},\ \bibinfo {pages} {245432} (\bibinfo
  {year} {2015}{\natexlab{a}})}\BibitemShut {NoStop}%
\bibitem [{\citenamefont {Le~Hur}(2012)}]{LeHur.12}%
  \BibitemOpen
  \bibfield  {author} {\bibinfo {author} {\bibfnamefont {K.}~\bibnamefont
  {Le~Hur}},\ }\href@noop {} {\bibfield  {journal} {\bibinfo  {journal} {Phys.
  Rev. B}\ }\textbf {\bibinfo {volume} {85}},\ \bibinfo {pages} {140506}
  (\bibinfo {year} {2012})}\BibitemShut {NoStop}%
\bibitem [{\citenamefont {Dmytruk}\ \emph
  {et~al.}(2015{\natexlab{b}})\citenamefont {Dmytruk}, \citenamefont {Trif},
  \citenamefont {Mora},\ and\ \citenamefont {Simon}}]{Dmytruk.15b}%
  \BibitemOpen
  \bibfield  {author} {\bibinfo {author} {\bibfnamefont {O.}~\bibnamefont
  {Dmytruk}}, \bibinfo {author} {\bibfnamefont {M.}~\bibnamefont {Trif}},
  \bibinfo {author} {\bibfnamefont {C.}~\bibnamefont {Mora}}, \ and\ \bibinfo
  {author} {\bibfnamefont {P.}~\bibnamefont {Simon}},\ }\href@noop {}
  {\bibfield  {journal} {\bibinfo  {journal} {arXiv:1510.03748v2}\ } (\bibinfo
  {year} {2015}{\natexlab{b}})}\BibitemShut {NoStop}%
\bibitem [{\citenamefont {Schir\'o}\ and\ \citenamefont
  {Le~Hur}(2014)}]{Schiro.14}%
  \BibitemOpen
  \bibfield  {author} {\bibinfo {author} {\bibfnamefont {M.}~\bibnamefont
  {Schir\'o}}\ and\ \bibinfo {author} {\bibfnamefont {K.}~\bibnamefont
  {Le~Hur}},\ }\href@noop {} {\bibfield  {journal} {\bibinfo  {journal} {Phys.
  Rev. B}\ }\textbf {\bibinfo {volume} {89}},\ \bibinfo {pages} {195127}
  (\bibinfo {year} {2014})}\BibitemShut {NoStop}%
\bibitem [{\citenamefont {Gudmundsson}\ \emph {et~al.}(2012)\citenamefont
  {Gudmundsson}, \citenamefont {Jonasson}, \citenamefont {Tang}, \citenamefont
  {Goan},\ and\ \citenamefont {Manolescu}}]{Gudmundsson.12}%
  \BibitemOpen
  \bibfield  {author} {\bibinfo {author} {\bibfnamefont {V.}~\bibnamefont
  {Gudmundsson}}, \bibinfo {author} {\bibfnamefont {O.}~\bibnamefont
  {Jonasson}}, \bibinfo {author} {\bibfnamefont {C.-S.}\ \bibnamefont {Tang}},
  \bibinfo {author} {\bibfnamefont {H.-S.}\ \bibnamefont {Goan}}, \ and\
  \bibinfo {author} {\bibfnamefont {A.}~\bibnamefont {Manolescu}},\ }\href@noop
  {} {\bibfield  {journal} {\bibinfo  {journal} {Phys. Rev. B}\ }\textbf
  {\bibinfo {volume} {85}},\ \bibinfo {pages} {075306} (\bibinfo {year}
  {2012})}\BibitemShut {NoStop}%
\bibitem [{\citenamefont {Arnold}\ \emph {et~al.}(2013)\citenamefont {Arnold},
  \citenamefont {Tang}, \citenamefont {Manolescu},\ and\ \citenamefont
  {Gudmundsson}}]{Arnold.13}%
  \BibitemOpen
  \bibfield  {author} {\bibinfo {author} {\bibfnamefont {T.}~\bibnamefont
  {Arnold}}, \bibinfo {author} {\bibfnamefont {C.-S.}\ \bibnamefont {Tang}},
  \bibinfo {author} {\bibfnamefont {A.}~\bibnamefont {Manolescu}}, \ and\
  \bibinfo {author} {\bibfnamefont {V.}~\bibnamefont {Gudmundsson}},\
  }\href@noop {} {\bibfield  {journal} {\bibinfo  {journal} {Phys. Rev. B}\
  }\textbf {\bibinfo {volume} {87}},\ \bibinfo {pages} {035314} (\bibinfo
  {year} {2013})}\BibitemShut {NoStop}%
\bibitem [{\citenamefont {Shiba}(1968)}]{Shiba.68}%
  \BibitemOpen
  \bibfield  {author} {\bibinfo {author} {\bibfnamefont {H.}~\bibnamefont
  {Shiba}},\ }\href@noop {} {\bibfield  {journal} {\bibinfo  {journal}
  {Progress of theoretical Physics}\ }\textbf {\bibinfo {volume} {40}},\
  \bibinfo {pages} {435} (\bibinfo {year} {1968})}\BibitemShut {NoStop}%
\bibitem [{\citenamefont {Yao}\ \emph {et~al.}(2014)\citenamefont {Yao},
  \citenamefont {Moca}, \citenamefont {Weymann}, \citenamefont {Sau},
  \citenamefont {Lukin}, \citenamefont {Demler},\ and\ \citenamefont
  {Zar\'and}}]{Yao.14}%
  \BibitemOpen
  \bibfield  {author} {\bibinfo {author} {\bibfnamefont {N.~Y.}\ \bibnamefont
  {Yao}}, \bibinfo {author} {\bibfnamefont {C.~P.}\ \bibnamefont {Moca}},
  \bibinfo {author} {\bibfnamefont {I.}~\bibnamefont {Weymann}}, \bibinfo
  {author} {\bibfnamefont {J.~D.}\ \bibnamefont {Sau}}, \bibinfo {author}
  {\bibfnamefont {M.~D.}\ \bibnamefont {Lukin}}, \bibinfo {author}
  {\bibfnamefont {E.~A.}\ \bibnamefont {Demler}}, \ and\ \bibinfo {author}
  {\bibfnamefont {G.}~\bibnamefont {Zar\'and}},\ }\href@noop {} {\bibfield
  {journal} {\bibinfo  {journal} {Phys. Rev. B}\ }\textbf {\bibinfo {volume}
  {90}},\ \bibinfo {pages} {241108} (\bibinfo {year} {2014})}\BibitemShut
  {NoStop}%
\bibitem [{\citenamefont {Mendes}\ and\ \citenamefont
  {Mora}(2015)}]{Mendes.15}%
  \BibitemOpen
  \bibfield  {author} {\bibinfo {author} {\bibfnamefont {U.~C.}\ \bibnamefont
  {Mendes}}\ and\ \bibinfo {author} {\bibfnamefont {C.}~\bibnamefont {Mora}},\
  }\href@noop {} {\bibfield  {journal} {\bibinfo  {journal} {New Journal of
  Physics}\ }\textbf {\bibinfo {volume} {17}},\ \bibinfo {pages} {113014}
  (\bibinfo {year} {2015})}\BibitemShut {NoStop}%
\bibitem [{\citenamefont {Wilson}(1975)}]{Wilson.75}%
  \BibitemOpen
  \bibfield  {author} {\bibinfo {author} {\bibfnamefont {K.~G.}\ \bibnamefont
  {Wilson}},\ }\href@noop {} {\bibfield  {journal} {\bibinfo  {journal} {Rev.
  Mod. Phys.}\ }\textbf {\bibinfo {volume} {47}},\ \bibinfo {pages} {773}
  (\bibinfo {year} {1975})}\BibitemShut {NoStop}%
\bibitem [{\citenamefont {Krishna-murthy}\ \emph {et~al.}(1980)\citenamefont
  {Krishna-murthy}, \citenamefont {Wilkins},\ and\ \citenamefont
  {Wilson}}]{Wilson.80}%
  \BibitemOpen
  \bibfield  {author} {\bibinfo {author} {\bibfnamefont {H.~R.}\ \bibnamefont
  {Krishna-murthy}}, \bibinfo {author} {\bibfnamefont {J.~W.}\ \bibnamefont
  {Wilkins}}, \ and\ \bibinfo {author} {\bibfnamefont {K.~G.}\ \bibnamefont
  {Wilson}},\ }\href@noop {} {\bibfield  {journal} {\bibinfo  {journal} {Phys.
  Rev. B}\ }\textbf {\bibinfo {volume} {21}},\ \bibinfo {pages} {1003}
  (\bibinfo {year} {1980})}\BibitemShut {NoStop}%
\bibitem [{\citenamefont {Hewson}\ and\ \citenamefont
  {Meyer}(2002)}]{Hewson.02}%
  \BibitemOpen
  \bibfield  {author} {\bibinfo {author} {\bibfnamefont {A.~C.}\ \bibnamefont
  {Hewson}}\ and\ \bibinfo {author} {\bibfnamefont {D.}~\bibnamefont {Meyer}},\
  }\href@noop {} {\bibfield  {journal} {\bibinfo  {journal} {Journal of
  Physics: Condensed Matter}\ }\textbf {\bibinfo {volume} {14}},\ \bibinfo
  {pages} {427} (\bibinfo {year} {2002})}\BibitemShut {NoStop}%
\bibitem [{\citenamefont {Jeon}\ \emph {et~al.}(2003)\citenamefont {Jeon},
  \citenamefont {Park},\ and\ \citenamefont {Choi}}]{Jeon.03}%
  \BibitemOpen
  \bibfield  {author} {\bibinfo {author} {\bibfnamefont {G.~S.}\ \bibnamefont
  {Jeon}}, \bibinfo {author} {\bibfnamefont {T.-H.}\ \bibnamefont {Park}}, \
  and\ \bibinfo {author} {\bibfnamefont {H.-Y.}\ \bibnamefont {Choi}},\
  }\href@noop {} {\bibfield  {journal} {\bibinfo  {journal} {Phys. Rev. B}\
  }\textbf {\bibinfo {volume} {68}},\ \bibinfo {pages} {045106} (\bibinfo
  {year} {2003})}\BibitemShut {NoStop}%
\bibitem [{hew()}]{hewson}%
  \BibitemOpen
  \href@noop {} {\bibinfo  {journal} {A. C. Hewson, {\it The Kondo Problem to
  Heavy Fermions} (Cambridge University Press, Cambridge, 1993)}\ }\BibitemShut
  {NoStop}%
\bibitem [{\citenamefont {Rozhkov}\ and\ \citenamefont
  {Arovas}(1999)}]{Rozhkov.99}%
  \BibitemOpen
\bibfield  {journal} {  }\bibfield  {author} {\bibinfo {author} {\bibfnamefont
  {A.~V.}\ \bibnamefont {Rozhkov}}\ and\ \bibinfo {author} {\bibfnamefont
  {D.~P.}\ \bibnamefont {Arovas}},\ }\href@noop {} {\bibfield  {journal}
  {\bibinfo  {journal} {Phys. Rev. Lett.}\ }\textbf {\bibinfo {volume} {82}},\
  \bibinfo {pages} {2788} (\bibinfo {year} {1999})}\BibitemShut {NoStop}%
\bibitem [{\citenamefont {Zonda}\ \emph {et~al.}(2015)\citenamefont {Zonda},
  \citenamefont {Pokorny}, \citenamefont {Janis},\ and\ \citenamefont
  {Novotny}}]{Zonda.15}%
  \BibitemOpen
  \bibfield  {author} {\bibinfo {author} {\bibfnamefont {M.}~\bibnamefont
  {Zonda}}, \bibinfo {author} {\bibfnamefont {V.}~\bibnamefont {Pokorny}},
  \bibinfo {author} {\bibfnamefont {V.}~\bibnamefont {Janis}}, \ and\ \bibinfo
  {author} {\bibfnamefont {T.}~\bibnamefont {Novotny}},\ }\href@noop {}
  {\bibfield  {journal} {\bibinfo  {journal} {Scientific Reports}\ }\textbf
  {\bibinfo {volume} {5}},\ \bibinfo {pages} {8821} (\bibinfo {year}
  {2015})}\BibitemShut {NoStop}%
\bibitem [{\citenamefont {Hecht}\ \emph {et~al.}(2008)\citenamefont {Hecht},
  \citenamefont {Weichselbaum}, \citenamefont {von Delft},\ and\ \citenamefont
  {Bulla}}]{Hecht.08}%
  \BibitemOpen
  \bibfield  {author} {\bibinfo {author} {\bibfnamefont {T.}~\bibnamefont
  {Hecht}}, \bibinfo {author} {\bibfnamefont {A.}~\bibnamefont {Weichselbaum}},
  \bibinfo {author} {\bibfnamefont {J.}~\bibnamefont {von Delft}}, \ and\
  \bibinfo {author} {\bibfnamefont {R.}~\bibnamefont {Bulla}},\ }\href@noop {}
  {\bibfield  {journal} {\bibinfo  {journal} {Journal of Physics: Condensed
  Matter}\ }\textbf {\bibinfo {volume} {20}},\ \bibinfo {pages} {275213}
  (\bibinfo {year} {2008})}\BibitemShut {NoStop}%
\bibitem [{\citenamefont {Meng}\ \emph {et~al.}(2009)\citenamefont {Meng},
  \citenamefont {Florens},\ and\ \citenamefont {Simon}}]{Meng.09}%
  \BibitemOpen
  \bibfield  {author} {\bibinfo {author} {\bibfnamefont {T.}~\bibnamefont
  {Meng}}, \bibinfo {author} {\bibfnamefont {S.}~\bibnamefont {Florens}}, \
  and\ \bibinfo {author} {\bibfnamefont {P.}~\bibnamefont {Simon}},\
  }\href@noop {} {\bibfield  {journal} {\bibinfo  {journal} {Phys. Rev. B}\
  }\textbf {\bibinfo {volume} {79}},\ \bibinfo {pages} {224521} (\bibinfo
  {year} {2009})}\BibitemShut {NoStop}%
\bibitem [{\citenamefont {Jaynes}\ and\ \citenamefont
  {Cummings}(1963)}]{Cummings.63}%
  \BibitemOpen
  \bibfield  {author} {\bibinfo {author} {\bibfnamefont {E.}~\bibnamefont
  {Jaynes}}\ and\ \bibinfo {author} {\bibfnamefont {F.}~\bibnamefont
  {Cummings}},\ }\href@noop {} {\bibfield  {journal} {\bibinfo  {journal}
  {Proceedings of the IEEE}\ }\textbf {\bibinfo {volume} {51}},\ \bibinfo
  {pages} {89} (\bibinfo {year} {1963})}\BibitemShut {NoStop}%
\bibitem [{Man()}]{Mandel}%
  \BibitemOpen
  \href@noop {} {\bibinfo  {journal} {Leonard Mandel, Emil Wolf {\it Optical
  Coherence and Quantum Optics} (Cambridge University Press, 1995)}\
  }\BibitemShut {NoStop}%
\bibitem [{\citenamefont {Deng}\ \emph
  {et~al.}(2015{\natexlab{b}})\citenamefont {Deng}, \citenamefont {Wei},
  \citenamefont {Johansson}, \citenamefont {Zhang}, \citenamefont {Li},
  \citenamefont {Li}, \citenamefont {Cao}, \citenamefont {Xiao}, \citenamefont
  {Tu}, \citenamefont {Guo}, \citenamefont {Jiang}, \citenamefont {Nori},\ and\
  \citenamefont {Guo}}]{Deng.15b}%
  \BibitemOpen
\bibfield  {journal} {  }\bibfield  {author} {\bibinfo {author} {\bibfnamefont
  {G.-W.}\ \bibnamefont {Deng}}, \bibinfo {author} {\bibfnamefont
  {D.}~\bibnamefont {Wei}}, \bibinfo {author} {\bibfnamefont {J.~R.}\
  \bibnamefont {Johansson}}, \bibinfo {author} {\bibfnamefont {M.-L.}\
  \bibnamefont {Zhang}}, \bibinfo {author} {\bibfnamefont {S.-X.}\ \bibnamefont
  {Li}}, \bibinfo {author} {\bibfnamefont {H.-O.}\ \bibnamefont {Li}}, \bibinfo
  {author} {\bibfnamefont {G.}~\bibnamefont {Cao}}, \bibinfo {author}
  {\bibfnamefont {M.}~\bibnamefont {Xiao}}, \bibinfo {author} {\bibfnamefont
  {T.}~\bibnamefont {Tu}}, \bibinfo {author} {\bibfnamefont {G.-C.}\
  \bibnamefont {Guo}}, \bibinfo {author} {\bibfnamefont {H.-W.}\ \bibnamefont
  {Jiang}}, \bibinfo {author} {\bibfnamefont {F.}~\bibnamefont {Nori}}, \ and\
  \bibinfo {author} {\bibfnamefont {G.-P.}\ \bibnamefont {Guo}},\ }\href@noop
  {} {\bibfield  {journal} {\bibinfo  {journal} {Phys. Rev. Lett.}\ }\textbf
  {\bibinfo {volume} {115}},\ \bibinfo {pages} {126804} (\bibinfo {year}
  {2015}{\natexlab{b}})}\BibitemShut {NoStop}%
\bibitem [{\citenamefont {Gasse}\ \emph {et~al.}(2013)\citenamefont {Gasse},
  \citenamefont {Lupien},\ and\ \citenamefont {Reulet}}]{Gasse.13}%
  \BibitemOpen
  \bibfield  {author} {\bibinfo {author} {\bibfnamefont {G.}~\bibnamefont
  {Gasse}}, \bibinfo {author} {\bibfnamefont {C.}~\bibnamefont {Lupien}}, \
  and\ \bibinfo {author} {\bibfnamefont {B.}~\bibnamefont {Reulet}},\
  }\href@noop {} {\bibfield  {journal} {\bibinfo  {journal} {Phys. Rev. Lett.}\
  }\textbf {\bibinfo {volume} {111}},\ \bibinfo {pages} {136601} (\bibinfo
  {year} {2013})}\BibitemShut {NoStop}%
\bibitem [{\citenamefont {Bulla}\ \emph {et~al.}(1998)\citenamefont {Bulla},
  \citenamefont {Hewson},\ and\ \citenamefont {Pruschke}}]{Bulla.98}%
  \BibitemOpen
  \bibfield  {author} {\bibinfo {author} {\bibfnamefont {R.}~\bibnamefont
  {Bulla}}, \bibinfo {author} {\bibfnamefont {A.~C.}\ \bibnamefont {Hewson}}, \
  and\ \bibinfo {author} {\bibfnamefont {T.}~\bibnamefont {Pruschke}},\
  }\href@noop {} {\bibfield  {journal} {\bibinfo  {journal} {Journal of
  Physics: Condensed Matter}\ }\textbf {\bibinfo {volume} {10}},\ \bibinfo
  {pages} {8365} (\bibinfo {year} {1998})}\BibitemShut {NoStop}%
\bibitem [{\citenamefont {Gaass}\ \emph {et~al.}(2014)\citenamefont {Gaass},
  \citenamefont {Pfaller}, \citenamefont {Geiger}, \citenamefont {Donarini},
  \citenamefont {Grifoni}, \citenamefont {H\"uttel},\ and\ \citenamefont
  {Strunk}}]{Gaass.14}%
  \BibitemOpen
  \bibfield  {author} {\bibinfo {author} {\bibfnamefont {M.}~\bibnamefont
  {Gaass}}, \bibinfo {author} {\bibfnamefont {S.}~\bibnamefont {Pfaller}},
  \bibinfo {author} {\bibfnamefont {T.}~\bibnamefont {Geiger}}, \bibinfo
  {author} {\bibfnamefont {A.}~\bibnamefont {Donarini}}, \bibinfo {author}
  {\bibfnamefont {M.}~\bibnamefont {Grifoni}}, \bibinfo {author} {\bibfnamefont
  {A.~K.}\ \bibnamefont {H\"uttel}}, \ and\ \bibinfo {author} {\bibfnamefont
  {C.}~\bibnamefont {Strunk}},\ }\href@noop {} {\bibfield  {journal} {\bibinfo
  {journal} {Phys. Rev. B}\ }\textbf {\bibinfo {volume} {89}},\ \bibinfo
  {pages} {241405} (\bibinfo {year} {2014})}\BibitemShut {NoStop}%
\bibitem [{Neg()}]{Negele}%
  \BibitemOpen
  \href@noop {} {\bibinfo  {journal} {John W. Negele, {\it Quantum
  Many-particle Systems} (Westview Press, 1998)}\ }\BibitemShut {NoStop}%
\end{thebibliography}%

\end{document}